\documentclass{mn2e}
\usepackage{psfig}
\usepackage{times}
\usepackage{amsmath}
\usepackage{multicol}
\usepackage[dvipsnames,usenames]{color}
\usepackage{colortbl}
\newcommand{\etal}{et al.\ }
\newcommand{\new}[1]{{\bf\textcolor{Red}{#1}}}

\newif\ifAMStwofonts

\newcommand{\lapp}{\la}
\newcommand{\gapp}{\ga}

\title[Double-double radio galaxies]{Particle
  acceleration and dynamics of double-double radio galaxies: theory vs. observations}
\author[C. Konar \& M.J. Hardcastle]
       {C. Konar$^1$ $\thanks{E-mail: chiranjib.konar@gmail.com (CK)}$
         and M.J. Hardcastle$^2$ \\
$^1$ Institute of Astronomy and Astrophysics, Academia Sinica, National Taiwan University, Taipei 10617, Taiwan, R.O.C.  \\
$^2$ School of Physics, Astronomy and Mathematics, University of Hertfordshire, College Lane, Hatfield \\
}
\date{Accepted.    Received }

\pagerange{\pageref{firstpage}--\pageref{lastpage}}
\pubyear{2013}

\begin{document}
\maketitle

\label{firstpage}

\begin{abstract}
In this paper we show that a small sample of radio galaxies with
evidence for multiple epochs of jet activity (so-called
`double-double' radio galaxies) have the same electron injection
spectral index in the two activity episodes, a result which might be
considered surprising given the very different lobe dynamics expected
in the first and second episode. We construct models for the dynamics
of radio galaxies, with an emphasis on their episodic behaviour, and
show that hotspot formation and confinement of lobes for the inner
double of double-double radio galaxies are possible even without any
thermal matter in the outer cocoon. We argue that (i) the observed
similar injection spectral indices are due to similar jet powers in
the two episodes, (ii) the `spectral index--radio power' correlation
of a flux limited sample of radio galaxies is the primary one, and not
the `spectral index--redshift correlation', (iii) jets are made of
pair plasma and not electron-proton, (iv) and the Lorentz factor of
the spine of the jet should be $\gapp 10$ to explain the observations.
Furthermore, we argue that the observations show that higher power
radio galaxies do not have a higher jet bulk Lorentz factors, but
instead simply have a higher number density of particles in the jet
rest frame. A consequence of our models is that aligned double-double
radio galaxies with very old ($\gapp10^8$ yr) outer doubles, or
misaligned double-double radio galaxies, are statistically more likely
to have dissimilar injection indices in two different episodes, as
they will probably have different jet powers.
\end{abstract}


\section{Introduction}
\begin{figure*} 
\hbox{ 
 \psfig{file=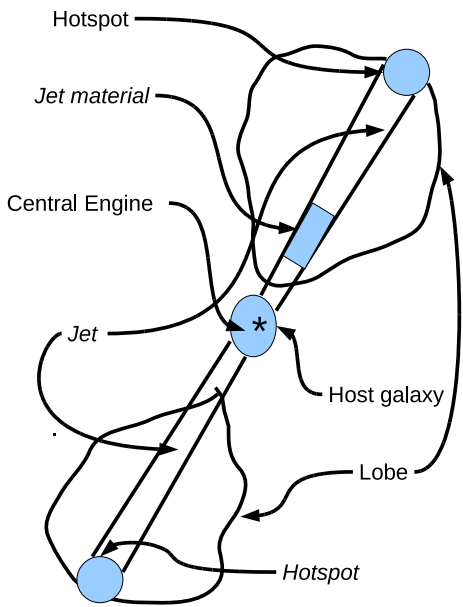,height=9.0cm,angle=0}
\hspace{1.5cm}
 \psfig{file=J1158+2621_GMRT-L.LR_color.red_v2.ps,height=8.8cm,angle=0}
}  
\caption[]{Left panel: A schematic picture of an FR\,II RG. Only a portion 
of the continuous jet-fluid is shown in blue. Right panel: A false colour 
image of a known episodic radio galaxy, J1158+2621 made with the GMRT at L band
(Konar et al., 2013).
}
\label{fig_schematics.n.real-RG}
\end{figure*}

Observations suggest (Richstone et al. 1998) that every massive galaxy
harbours a super-massive black hole (SMBH). Under favourable
conditions, the black hole and its surrounding accretion system can
launch relativistic jets carrying charged sub-atomic particles (e.g.,
e$^{-}$p$^{+}$ or e$^{-}$e$^{+}$) together with magnetic fields. These
jets supply particles and energy to lobes inflated by the jets
themselves, which may extend from kpc to Mpc scales. The complete
system, consisting of a SMBH at the centre of a galaxy, the jets, and
the lobes, is called a radio galaxy, and comparatively powerful jets
give rise to Fanaroff-Riley type II (FR\,II) radio galaxies (Fanaroff
\& Riley 1974). Observations have proven (e.g.\ Schoenmakers et
al.\ 2000) that the phenomenon of launching jets is episodic in nature
(see the right panel of Fig.~\ref{fig_schematics.n.real-RG} for images
of an episodic radio galaxy) with the duration of the quiescent phase
ranging from a tenth of a Myr to a few tens of Myr in several
well-studied cases (Konar et al.\ 2012, 2013). The speed of the jets
in FR\,II radio galaxies is known to be supersonic with respect to the
internal sound/magnetosonic speed of the lobe as well as the jet
fluid, and so the jet inevitably terminates in a shock called the
jet-termination shock (JTS). The observed bright radio features at the
end of the lobes, which are known as hotspots, are identified with the
JTS (see Fig.~\ref{fig_schematics.n.real-RG} for schematic picture of
a radio galaxy). At the JTS, some fraction of the kinetic energy of
the jet is put into a population of relativistic particles at the
shock by a process known as Diffusive Shock Acceleration (DSA, see
Bell 1978a, 1978b; Kirk et al. 2000) before the particles escape from
the shock/hotspot region and expand into the lobes. In the FRII radio
galaxies, the relativistic DSA gives rise to a power-law energy
distribution of the particles in the hotspots, which is given by $N(E)
\propto E^{-\delta}$, where $N(E)$ is the number of particles at
energy $E$ and $\delta$ is the power-law index. The leptons in the
jet, being lighter particles, are efficiently accelerated at the
hotspots and radiate via the synchrotron process; protons, being much
more massive, may still be accelerated but produce negligible amounts
of radiation by this mechanism. Since the energy distributions of the
relativistic particles are power laws, the synchrotron spectra from
regions of FR\,II RGs where particle acceleration has recently taken
place are also expected to be power laws with the general form
$S(\nu)=S_0\nu^{-\alpha_{\rm inj}}$, where $S(\nu)$ is the synchrotron
flux density from a given emission region of the lobes and/or
hotspots, $S_0$ is the normalization and $\alpha_{\rm inj}$ is the
injection spectral index (hereafter injection index). The index,
$\alpha_{\rm inj}$ is related to the power-law index of energy
distribution, $\delta$ by $\alpha_{\rm inj}=(\delta-1)/2$ (Pacholczyk
1980). In principle, the injection index $\alpha_{\rm inj}$ can give
us information about the properties of the JTS\footnote{Although
    observations of high-energy synchrotron radiation (e.g.
    Meisenheimer \etal\ 1989; Hardcastle \etal\ 2004) demonstrate that
    the hotspots are locations of high-energy particle acceleration,
    it is important to note that we cannot rule out the possibility of
    particle acceleration elsewhere in the sources. In particular,
    observations of emission from the jets, in some cases possibly
    extending to X-ray energies (e.g. Wilson \etal\ 2001), suggest
    that particles are (re)accelerated in the jet regions, before the
    JTS. The acceleration mechanism here is uncertain, but cannot be
    the result of jet-wide shocks stationary in the galaxy frame,
    which would make the downstream jet subsonic; some smaller-scale
    process must be responsible for the particle acceleration allowing
    emission to be seen from the jets. In this paper we make the
    assumption that the hotspots (the JTS), rather than the jets, are
    the location at which the bulk of the particle acceleration
    occurs, since it seems very likely that they are the location at
    which the bulk of the jet kinetic energy is thermalized; as we
    will argue later in the paper, a self-consistent model can be
    generated in which there is little bulk deceleration of the jet
    before the JTS. We are not aware of any direct method of testing
    this assumption.}. The actual observed spectra of large-scale
regions of these RGs are more complex, because of the effects of
radiative and inverse-Compton losses and adiabatic expansion, which
give rise to broken or curved synchrotron spectra (e.g. Heavens \&
Meisenheimer 1987, Jaffe \& Perola 1973) and so in general
measurements at several frequencies are needed to determine the
injection index.

The main objectives of the present paper are (i) to study observationally 
the behaviour of $\alpha_{\rm inj}$ in two different episodes of jet formation 
activity  of individual episodic radio galaxies, (ii) to study 
the dynamics of jets, especially the inner jets of episodic radio galaxies, 
and (iii) to try to determine the parameters of the system on which 
$\alpha_{\rm inj}$ depends. By doing so we can learn a good deal about 
the properties of radio galaxies in general.

  The remainder of the paper is structured as follows. In Section
  \ref{sample} we construct a small sample of episodic radio galaxies 
  from the literature and our own observations. Section \ref{results} 
  describes how we have measured the injection indices for the multiple 
  epochs of the sample of episodic radio galaxies, and how we estimate 
  jet power for these sources; we then present the key observational 
  results of this paper, including a strong correlation between the 
  injection indices seen in the two epochs of the episodic radio galaxy 
  activity. In Section \ref{sec_reldynamics.of.rg} we move on to discuss 
  relativistic models for radio galaxy dynamics, and their implications 
  for the expected properties of JTS and injection index in episodic radio 
  galaxies. Section \ref{sec_discussion} then discusses our observational 
  results in the context of these models. Our conclusions are given in 
  Section~\ref{sec_conclusion}.

\section{Sample compilation, observations and data reduction}
\label{sample}

We have compiled from the literature a sample of 8 known
double-double radio galaxies (DDRGs), which are listed in
Table~\ref{tab_alpha.corrln_inn.vs.out}. By definition, these objects
show two episodes of jet forming activity. Out of the 8 sample objects, 
we obtained new radio observations for 4 DDRGs (J0041+3224, J0116-4722, 
J1158+2621 and J1835+6204) with the GMRT at 150, 240, 332, 605 and 1287 
MHz and used archival VLA data at higher frequencies to estimate the 
injection indices. The observing logs, the method of observations and 
the data reduction for these objects are discussed in detail by Konar et al.
(2012; 2013). From the literature, we compiled flux
densities down to very low frequencies to constrain the radio spectra of
the outer doubles of these 4 DDRGs. The remaining 4 objects in our
sample already had published estimates of $\alpha_{\rm inj}$ for both
the inner and outer doubles, and we adopt those values in this paper.

\section{Spectra, injection indices and jet powers}
\label{results}
\subsection{Injection index}
\begin{table}
\caption{Injection index of inner and outer doubles for our DDRG
sample. Columns are as follows. Column 1: source name,
Column 2: $\alpha_{\rm inj}$ with error of the inner double, 
Column 3: $\alpha_{\rm inj}$ with error of the outer double, and 
Column 4: reference to the source of $\alpha_{\rm inj}$  and comment.}
\label{tab_alpha.corrln_inn.vs.out}
\begin{tabular}{l l l l l l r c r r c}
\hline
Source         & $\alpha^{\rm inn}_{\rm inj}$   & $\alpha^{\rm out}_{\rm inj}$   & Ref. and comment       \\
(1)            & (2)                    &  (3)                   & (4)                    \\
\hline
 J0041$+$3224             & 0.724$^{+0.034}_{-0.041}$         & 0.756$^{+0.167}_{-0.122}$              &   a      \\
 J0116$-$4722             & 0.700$^{+0.100}_{-0.100}$         & 0.618$^{+0.072}_{-0.065}$              &   a,b     \\ 
 J0840$+$2949             & 0.830$^{+0.050}_{-0.050}$         & 0.810$^{+0.050}_{-0.050}$              &   c      \\
 J1158$+$2621             & 0.768$^{+0.029}_{-0.047}$         & 0.788$^{+0.038}_{-0.040}$              &   a      \\
 J1352$+$3126             & 0.720$^{+0.020}_{-0.020}$         & 0.855$^{\ast}$$^{+0.030}_{-0.030}$     &   d       \\
 J1453$+$3308             & 0.566$^{+0.051}_{-0.058}$         & 0.568$^{+0.065}_{-0.060}$              &   e      \\
 J1548$-$3216             & 0.579$^{+0.113}_{-0.157}$         & 0.567$^{+0.070}_{-0.066}$              &   f      \\
 J1835$+$6204             & 0.860$^{+0.034}_{-0.488}$         & 0.818$^{+0.070}_{-0.064}$              &   a      \\
\hline
\end{tabular}
\begin{flushleft}
$^{\ast}$: The average $\alpha_{\rm inj}$ of two outer lobes from Joshi et al. (2011). \\
a: This paper for $\alpha_{\rm inj}$ of the inner double and Konar et al. (2012b, in prep) for $\alpha_{\rm inj}$ of the outer double.
b: Saripalli et al. (2002) for the injection index for the inner double. 
c: Jamrozy et al. (2007). d: Joshi et al. (2011). e : Konar et al. (2006).
f: We fitted the inner-double and outer-double spectra, using the data from the Table~2 of
   Machalski, jamrozy \& Konar (2010).
\end{flushleft}
\end{table}
In this section we summarize the method by which $\alpha_{\rm inj}$
was estimated from our new observations. For details of the
observations, see Konar et al.\ (2012, 2013).

The spectra of the inner doubles were constrained from the
high-frequency measurements. Most of the inner doubles of our DDRGs
are embedded in the diffuse emission of the relic outer lobes.
Obviously, the $uv$ data at different frequencies used for imaging
have different shortest baselines; so we re-mapped the fields of each
DDRG at higher frequencies with the same lower $uv$ cutoff in order to
image the inner doubles free from the outer diffuse emission. The
high-frequency flux densities and their errors, both for individual
components and for the total inner doubles of our sample DDRGs, were
measured from the maps re-made with similar lower uv cutoff; these
fluxes are tabulated by Konar et al. (2012; 2013). We assumed 7 per
cent flux density errors at 0.62 and 1.28 GHz and 5 per cent errors at
1.40, 4.86, 8.46 and 22.46 GHz for our flux density measurements of
each inner lobe. The errors on the total flux densities of the inner
doubles (without the core) were obtained by propagating the errors of
the two individual lobes. No appreciable curvature was visible in the
spectra of the individual inner lobes and the core-subtracted total
inner doubles of all the DDRGs within our observable frequency range.
We therefore constrained $\alpha_{\rm inj}$ of the inner doubles by
fitting power laws to their observed integrated spectra. For
J0116-4722, there are no high-resolution data at high frequencies
except for those at 1376 and 2496 MHz published by Saripalli et al.
(2002); for this source we have therefore used these two flux
densities to constrain the power-law spectrum of the inner double of
J0116-4722, on the assumption that the spectrum of the inner double of
this source is a power law within our observable range (10$-$22000
MHz). For the outer doubles, by contrast, the spectra have curvature
at higher frequencies, presumably due to synchrotron and
inverse-Compton losses. However, the low-frequency part of the spectra
are still power laws and have not yet been affected by the loss, so
that the spectral index of that part of the spectra is presumably a
good estimate of $\alpha_{\rm inj}$ (see Konar et al., 2006; 2012;
2013).

\subsection{Jet power}
\label{jetpower}
The particle acceleration in FRIIs is thought to be due to the JTS at the
hotspots. The efficiency of acceleration depends on the JTS strength
which is expected to depend on the jet power ($Q_{\rm j}$). The shock
strength depends on the upstream speed of the jet as observed from the 
hotspot frame (i.e. shock frame) which also moves with respect to 
the host galaxy frame. In a given ambient medium, the 
hotspot speed depends on the jet power and composition, through momentum balance at the 
contact discontinuity at the jet head. If the speed of the hotspot alters, 
the speed of the upstream jet with respect to the hotspot frame also alters 
(see Section~\ref{sec_mom.balance}). Hence, the shock strength and 
particle acceleration efficiency depends on the jet power, and so in
this section we consider methods for estimating this for our target sources.

 It is not in practice easy to estimate $Q_{\rm j}$ from observations
 of the jets themselves, and, even if it were, we would be unable to
 do so since jets are not easily detectable in our observations.
 Instead we estimate the jet power by measuring the total enthalpy of
 the lobes and dividing it by the spectral age of the source, assuming
 the minimum-energy condition. Mathematically, we can write $Q_{\rm j}
 = \frac{4PV}{t_{\rm activ}}$, where $P$ and $t_{\rm activ}$ are lobe
 pressure and the duration of active phase of the jets respectively at
 the minimum energy condition. $V$ is the lobe volume. To determine
 the spectral ages and hence $t_{\rm activ}$, we have fitted
 Jaffe-Perola models (Jaffe \& Perola 1973) to the radio spectra of
 all the sources. We used the formalism of Konar et al.\ (2008) to
 estimate the minimum energy magnetic field ($B_{\rm min}$), and hence
 to estimate the total pressure $P$ ($=\frac{1}{3}(\epsilon_{\rm
   B,min}+\epsilon_{\rm e,min}$)) of the lobes, which we take to be
 cylinders for the purposes of estimating volume. The method of
 estimating spectral age is described in detail by Konar et
 al.\ (2012, 2013). The method of estimating $Q_{\rm j}$ that we adopt
 neglects the work done on the external environment due to driving
 shocks into the external medium. However, as numerical modelling in
 realistic environments (Hardcastle \& Krause 2013) shows that the
 work done on the external medium is comparable to the internal energy
 of the lobes over the greater part of the lifetime of the RG, we
 expect this to give at worst a systematic underestimate of $Q_{\rm
   j}$, which would not affect any observed correlation between
 $Q_{\rm j}$ and $\alpha_{\rm inj}$.
\subsection{Correlations}
\label{sec_results}
\begin{figure}
\vbox{
 \psfig{file=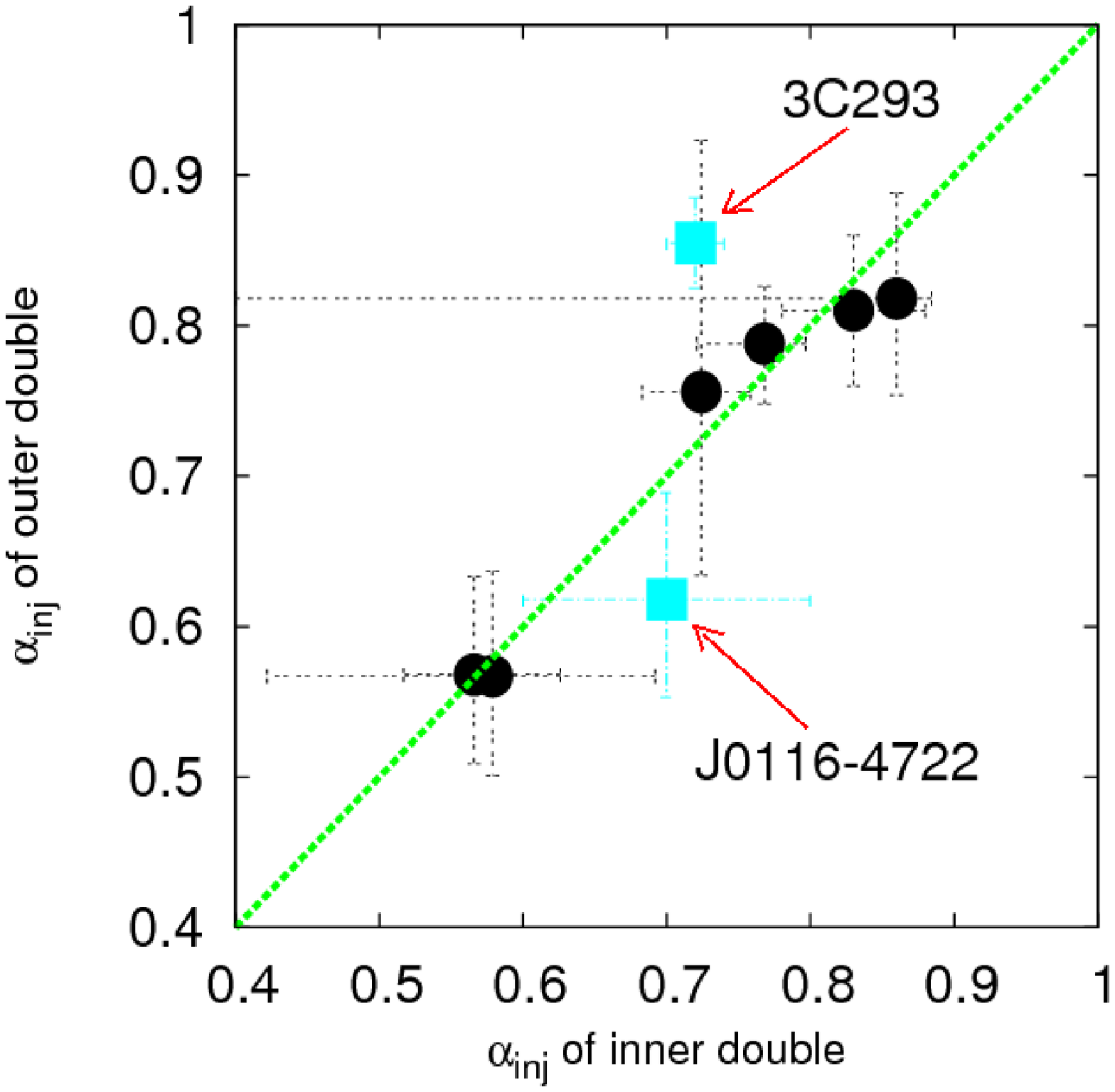,width=8.0cm,angle=0}
 \psfig{file=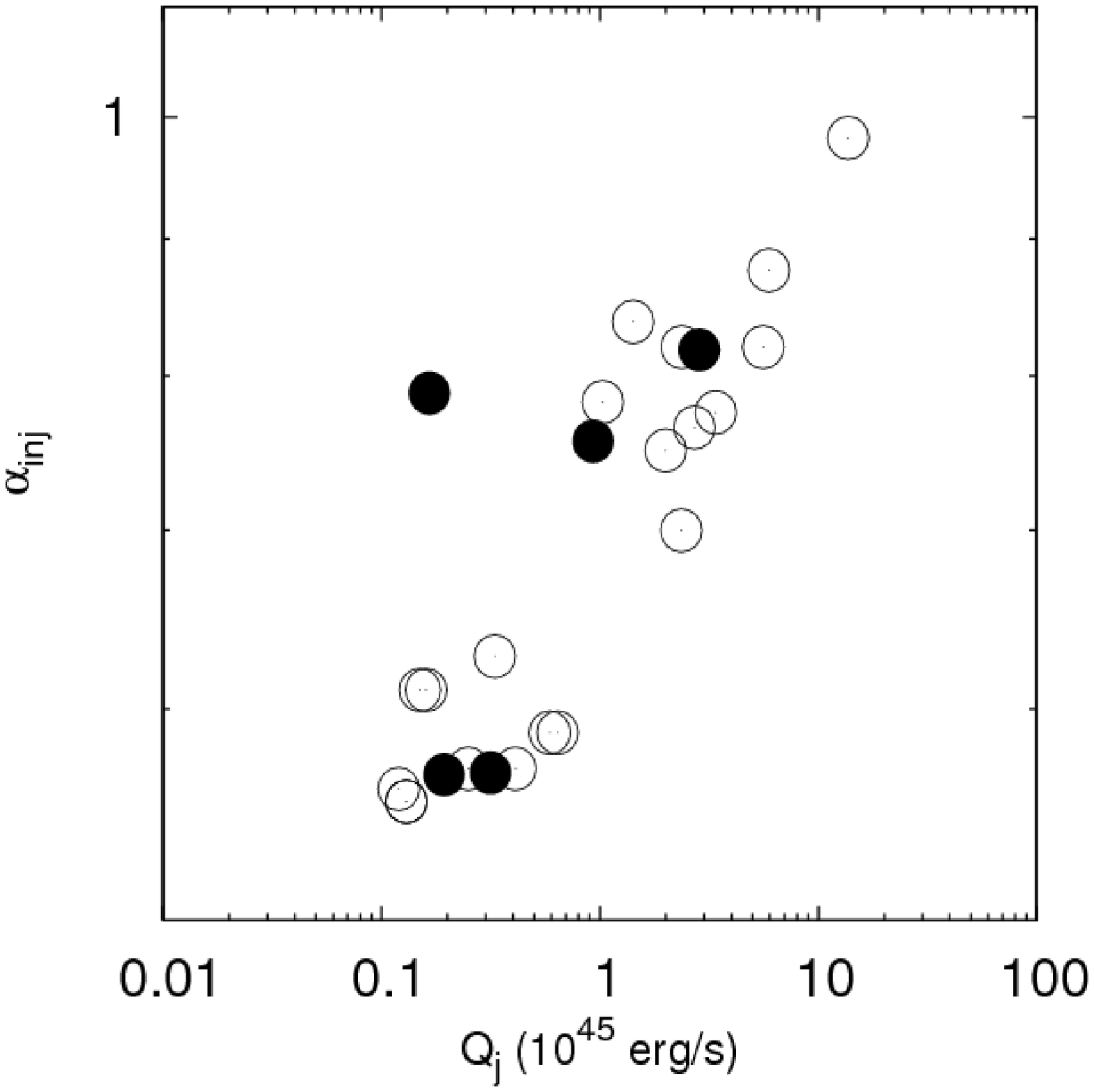,width=8.0cm,angle=0}
}
\caption[]{
Top panel: $\alpha_{\rm inj}$ for the outer and inner doubles of our
sample of DDRGs.
Two objects, indicated by filled triangles, have slightly dissimilar values 
of $\alpha_{\rm inj}$ in the two episodes, but all the other sources lie 
close to the line of equality (dotted line). Bottom panel: $\alpha_{\rm inj}$ 
vs. $Q_{\rm j}$ plot for the FRII RGs considered in this paper. A filled circle represents an entire 
outer double of a DDRG. An open circle represents a single lobe 
of a LRG, except 3C46 and 3C452 which are plotted as entire doubles.
}
\label{fig_alpha-alpha_corrln}
\end{figure}

In this section we discuss the results related to $\alpha_{\rm
inj}$ and $Q_{\rm j}$ for our sample. We begin by plotting 
$\alpha_{\rm inj}$ for the outer doubles against that of the inner 
doubles. This plot shows a correlation (see top panel of 
Fig.~\ref{fig_alpha-alpha_corrln}), significant at the 95 per cent 
level on a Spearman rank test, indicating that $\alpha_{\rm inj}$ 
is quite similar in the two different episodes of jet activity of 
most of the DDRGs in our sample. We discuss this remarkable result 
below.

We also attempted to find any possible correlation between 
$\alpha_{\rm inj}$ and $Q_{\rm j}$ of FR\,II radio galaxies. We compiled 
a small sample of Large Radio Galaxies (LRGs) which have well constrained 
spectra, and good estimation of $\alpha_{\rm inj}$ and spectral ages 
published by Jamrozy et al. (2008) and Nandi et al. (2010). From the 
spectra and ages, we estimated $Q_{\rm j}$ for our sample of DDRGs and 
LRGs (see Table~\ref{tab_su.z.alpha.jetpow}, online material), though we 
were unable to determine spectral ages for three of our DDRGs. The
determination of spectral ages, and the estimation of $\alpha_{\rm
  inj}$ and $Q_{\rm j}$, were carried out in the same way for both samples.
We plotted $\alpha_{\rm inj}$ vs. $Q_{\rm j}$ (see bottom panel of
Fig.~\ref{fig_alpha-alpha_corrln}) and found a correlation which is
significant at better than 99.9 per cent on a Spearman rank test.

To understand these results we need to consider the dynamics of radio
galaxies, particularly in the double-double phase, and that is the
topic of the following section of the paper.

\section{Relativistic dynamics of radio galaxies}
\label{sec_reldynamics.of.rg}

\subsection{Introduction}
Observationally it is clear that the formation of DDRG and
Triple-Double Radio Galaxy (TDRG) morphologies is due to the episodic
jet forming activity (see Konar et al. 2006, 2012, 2013). However,
there are two plausible models to explain how the inner double
morphology can be created. Those two models are the `classical FR\,II
model' and `bow shock model' (see Konar et al. 2013). In the first of
these, the inner lobes are formed in the same way as the Single-Double
Radio Galaxies (SDRGs) and the outer doubles of DDRGs, i.e., by the
back-flowing relativistic plasma injected at the hotspots. This is
what is referrred to as `classical FR\,II model'. In the `bow shock
model' the inner doubles are created by re-acceleration of the
particles at the bow shocks driven into the outer cocoon material
created by the almost ballistically moving jet heads (Brocksopp et al.
2007, 2011; Safouris et al. 2008). While some sources are indeed well
described by the `bow shock model', there is some morphological
evidence that it does not apply to all DDRG (Safouris \etal\ 2008). In
addition, crucially, there seems to be no real reason to expect a
similar injection index in the inner and outer doubles in this model,
as observed (Section \ref{sec_results}) since the shocks in the
hotspots of the outer doubles and the bow shock in the inner double
should have very different properties. Although the bow shock model
must be correct at some level (i.e. there must be some reacceleration
of particles if the new lobe drives a shock through the old one) the
observations of correlated injection indices suggest that emission
from the bow shock cannot dominate the observed inner
double\footnote{It is worth noting one special case: if the shock
  driven into the outer lobe is quite weak, so that there is little
  particle acceleration and instead simply adiabatic compression of
  the particles and field, then the compressed material would indeed
  be expected to have the same injection index as the outer lobe,
  since the inner double would simply be rejuvenated outer lobe
  material. This model may be applicable in some weak sources on
  galactic scales (see Mingo \etal\ 2012) but the other properties of
  the inner doubles do not seem consistent with weak shocks, so we do
  not consider it further here.}. In this paper we therefore
concentrate on the `classical FRII model'.

As discussed in more detail by Konar et al. (2013, Section 1), the
main problem with the classical FRII model for the inner doubles is,
or is perceived to be, the very rapid hotspot advance speed expected
due to the low density in the outer lobes. The presence of compact
components at the jet heads of the inner doubles of all FR\,II DDRGs,
together with the observation that the inner doubles can extend back
for significant distances towards the core, motivates the idea that
there is entrainment/ingestion of matter from the thermal ambient
medium into the large-scale lobes, increasing the internal density of
these lobes and hence reducing the hotspot advance speed. Kaiser,
Schoenmakers \& R{\"o}ttgering (2000) have argued that the
Kelvin-Helmholtz and Raleigh-Taylor instabilities do not grow quickly
in the contact discontinuity separating the magnetized ralativistic
plasma of the radio lobes and the thermal ambient medium. According to
them, the time scales ($\sim10^8$ yr) of those instabilities on the
relevant scale length ($\sim1$ kpc) are higher than the age of the
outer doubles of most of the known DDRGs. Kaiser et al. (2000) argued
that the thermal matter entrainment into the radio lobes is therefore
a slow and inefficient process. They proposed a model in which the
thermal matter is ingested into the cocoon (or lobes) in the following
way. When the jets propagate through the thermal ambient medium, the
jet-heads drives shocks into that ambient medium, which is a two phase
medium with cooler and denser clumps embedded in the hot gaseous
Inter-Galactic Medium (IGM). Since the clumps are heavier, they are
not imparted sufficient momentum by the passage of the bow shock. As a
result, the contact discontinuity overtakes those shocked clumps which
ends up being inside the cocoon, or the lobes. These clouds then
diffuse inside the cocoon throughout the cocoon volume, and finally
mix with the cocoon matter. This increases the mass density inside the
cocoon and provides a favourable situation for the inner double to form
hotspots and well-confined inner lobes. The main problem with this
model is the lack of any strong observational evidence that there is a
significant density of thermal matter in the lobes: this should be the
case for all large FRIIs, and so will be readily testable in coming
years with low-frequency polarization observations. In the meantime,
we shall explore the consequences of alternative models, in which we
assume that the FR\,II lobes contain only nonthermal pair plasma.

\subsection{Comparison of the dynamics of lobe protons, if at all present, with that of lobe electrons}
\label{sec_proton.dynamics}

Whether the jet matter consists of electron-proton (hereafre
$e^{-}p^{+}$) or electron-positron (hereafter $e^{-}e^{+}$) plasma is
so far a debatable issue. However, if the jet matter does consist
of $e^{-}p^{+}$ plasma, we would like to assess the possible
dynamical consequences of the additional heavy particles.

The results of Croston et al (2005) suggest that the mean kinetic
energy of lobe protons can be at most similar to the mean kinetic
energy of lobe electrons. So we can write
\[ \langle (\gamma_p - 1)m_pc^2\rangle < \langle (\gamma_e - 1) m_ec^2 \rangle \]
where $\gamma_p$ and $\gamma_e$ are the Lorentz factors of a proton and an electron respectively, 
$m_p$ and $m_e$ are the masses of a proton and an electron respectively and $c$ is the speed of light.
The angular brackets represent the mean of a quantity.
After simplifying the above relation we get
\begin{equation}
\langle \gamma_p \rangle  <  1 + (\langle \gamma_e\rangle - 1) \left(\frac{m_e}{m_p}\right)
\label{eqn_kin.energy_1}
\end{equation}
The mean Lorentz factor of the radiating particles in a radio lobe can be given by
\[ \langle \gamma_e \rangle =\frac{ \int \gamma_e N(\gamma_e) d\gamma_e }{\int N(\gamma_e) d\gamma_e}. \]
Assuming a power law distribution of radiating electrons,
$N(\gamma_e)=N_0\gamma_e^{-\delta}$ ($\delta > 2$) and 
integrating from $\gamma_1$ to $\gamma_2$, we obtain
\begin{equation}
\langle \gamma_e \rangle = \frac{\delta -1}{\delta -2}\frac{\gamma_1^{2-\delta}-\gamma_2^{2-\delta}}
                                          {\gamma_1^{1-\delta}-\gamma_2^{1-\delta} }
\label{eqn_mean_gamma_e}
\end{equation}
For FR\,II radio lobes the electrons have Lorentz factors from 10 to $>10^5$ with a typical
power law index of 2.5. For our representative estimate we assume that $\gamma_1=10$,
$\gamma_2 \rightarrow \infty$ and $\frac{m_e}{m_p}=\frac{1}{1836}$. This yields 
$\langle \gamma_e \rangle \sim 30$. If we estimate the average value of 
$\langle \gamma_p \rangle$ from equation~(\ref{eqn_kin.energy_1}) of such typical FR\,II 
lobes, we obtain 
\begin{equation}
1 \la \langle \gamma_p \rangle < 1.0158
\label{eqn_avg.gamma_p}
\end{equation}
(as Lorent factor cannot be lower than 1). A Lorentz factor of 1.0158
corresponds to a speed of $0.1757c$, which is mildly relativistic. We
may define an effective temperature ($T^{\rm L}_p$) for the lobe protons such that
\begin{equation}
 k_BT^{\rm L}_p = (\langle \gamma_p \rangle -1)m_pc^2, 
\end{equation}
where $k_B$ is the Boltzmann constant. This gives an average effective
temperature $T^{\rm L}_p < 1.72\times 10^{11}$ K, which is similar to
the nonthermal temperature (see equations~(\ref{eqn_pressure_nt_e})
and (\ref{eqn_int.energy_nt_e}) for definition) of radiating
particles. This simply means that the protons, if at all fed by the
FR\,II jets, have not been accelerated in the JTS as efficiently as
the electrons. In the relativistic DSA model, the flattest value of
$\alpha_{\rm inj} \sim 0.62$ (Kirk et al. 2000). For $\alpha_{\rm inj}
\sim 0.62$, we will get the maximum possible value of $\langle
\gamma_e \rangle$, $\langle \gamma_p \rangle$ and $T^{\rm L}_p$. Those
values are $\langle \gamma_e \rangle = 51.67 $, $\langle \gamma_p
\rangle < 1.0276$ and $T^{\rm L}_p < 3.01\times 10^{11}$ K.

Since we have an estimate of $\langle \gamma_e \rangle$, we can
estimate the number density of radiating particles in the radio lobes
in the following way. The radiating particles are in energy
equipartition with the magnetic field, so we can write
\[ (\langle \gamma_e \rangle -1) n_e m_ec^2 = \frac{B_{\rm eq}^2}{8\pi}, \]
where $B_{\rm eq}$ is the equipartition magnetic field strength.
From this we get an expression for the number density of electrons in the lobes:
\begin{equation}
n_e =  \frac{1}{(\langle \gamma_e \rangle -1)m_ec^2} \frac{B_{\rm eq}^2}{8\pi}.  
\label{eqn_estimate.n_e}
\end{equation} 
For $B_{\rm eq}=5$ $\mu$G, $\gamma_1 = 10$ and $\gamma_2 \rightarrow \infty$ limit, 
which are typical for the outer lobes of DDRGs,
equations~(\ref{eqn_mean_gamma_e}) and (\ref{eqn_estimate.n_e}) yield
\[ n_e \sim 7\times 10^{-8} {\rm cm}^{-3}\]  for $\alpha_{\rm inj}=0.62$ (flattest possible value)
and 
\[ n_e \sim 4\times 10^{-8} {\rm cm}^{-3}\] for $\alpha_{\rm inj}=0.75$ (a typical value).
Thus, if the jets are $e^{-}p^{+}$, an equal number of protons
in the outer lobe plasma would be more than enough to allow
hotspot formation in the inner jet head and confinement of the inner
lobes, a point that we shall return to in Section~\ref{sec_JTS.n.bowshock}.
(However, we shall also show that hotspot formation and inner lobe
confinement are possible even for a pure $e^{-}e^{+}$ plasma in the outer
cocoon.)

As a summary of this section, we can write  
\begin{equation}
\langle E^{\rm kin}_p \rangle  < \langle E^{\rm kin}_e \rangle,
\label{eqn_Ep.lt.Ee}
\end{equation}
and 
\begin{equation}
1 \lapp \langle \gamma_p \rangle < 1.0276,
\label{eqn_gamma_p_limit}
\end{equation}

These place very good constraints on the proton acceleration at the
JTS, if at all protons exist in the jet material. In the next section
we show that this implies that it is highly likely that the jet
composition is $e^{-}e^{+}$.

\subsection{Composition of FR\,II jets}
Here we argue that $e^-e^+$ jets are more viable than $e^-p^+$ jets,
at least in FR\,II radio galaxies. If the FR\,II jets are made of
$e^-p^+$, then in the jet flow electrons and protons must travel
together with the bulk Lorentz factor, $\Gamma_{\rm j}$ of the jet to
avoid charge separation. The constancy of spectral indices in two
episodes of DDRGs demands that the $\Gamma_{\rm j}$ be $>10$ (see
Section~\ref{sec_jp-index}), so that the JTS can be considered to be a
strong shock. Therefore, even if the protons collectively behave as a
(probably collisionless) background fluid in which the electrons are
embedded as test particles, a significant amount of the bulk kinetic
energy of the upstream flow is expected to be converted into the
internal energy of the proton fluid. Since the JTS is a strong
relativistic shock, the average value of the kinetic energy of the
protons would be expected to be {\it higher} than that of the leptons
in the lobes. As Drury (1983) suggests, ``The abundances of a species
in the high energy particles relative to that in the (upstream) plasma
should be a smooth and probably increasing function of its mass to
charge ratio". This is the so-called `selectivity of injection' of the
particles, though there exists no quantitative theory for it.
Effectively, protons in the upstream jet fluids will be more easily
injected into the shock to participate in the DSA than the electrons:
this will give rise to a larger number of accelerated protons than
leptons in the lobes. Though Drury advanced the concept of
`selectivity of injection' in the context of non-relativistic DSA,
there is no physically plausible reason why this `selectivity of
injection' should not be qualititatively valid for relativistic DSA as
well. However, as shown in the previous subsection, the results of
Croston et al. (2005) indicate that the protons are energetically not
dominant, and thus cannot be accelerated to higher energies than electrons.
Hence, if our assumptions hold, jets must be composed of $e^-e^+$
rather than $e^-p^+$. We will consider the jet plasma to be $e^-e^+$
in what follows. 

  We note that work on FR\,I radio galaxies (e.g. McNamara \&
  Nulsen, 2012) suggests that there are protons inside the FR\,I
  lobes. Recent observational results suggest that this is most
  probably due to the entrainment of thermal matter through the FR\,I
  jets (Croston et al. 2008), while there are various lines of
  evidence (e.g. Dunn, Fabian \& Celotti 2006) that FR\,I jets are
  electron-positron at their bases. We would argue therefore that jets
  are originally $e^-e^+$ for both FR\,I and FR\,II sources.
  Entrainment both slows down the FR\,I jets and causes their lobes to
  have a non-negligible proton content. FR\,II jets are not so
  strongly affected by entrainment (Bicknell 1995) and their lobes
  should thus be mainly filled with $e^-e^+$ plasma, consistent with
  observations (e.g. Croston \etal\ 2004) that show that pressure
  balance can be achieved in the lobes with only $e^-e^+$ together
  with magnetic fields whose strengths are derived from
  inverse-Compton observations. Since our DDRGs are all FR\,IIs, we
  will consider both jet and lobe plasma to be $e^-e^+$ in what
  follows.

\subsection{Momentum balance at the jet heads}
\label{sec_mom.balance}
\begin{figure}
\hbox{
 \psfig{file=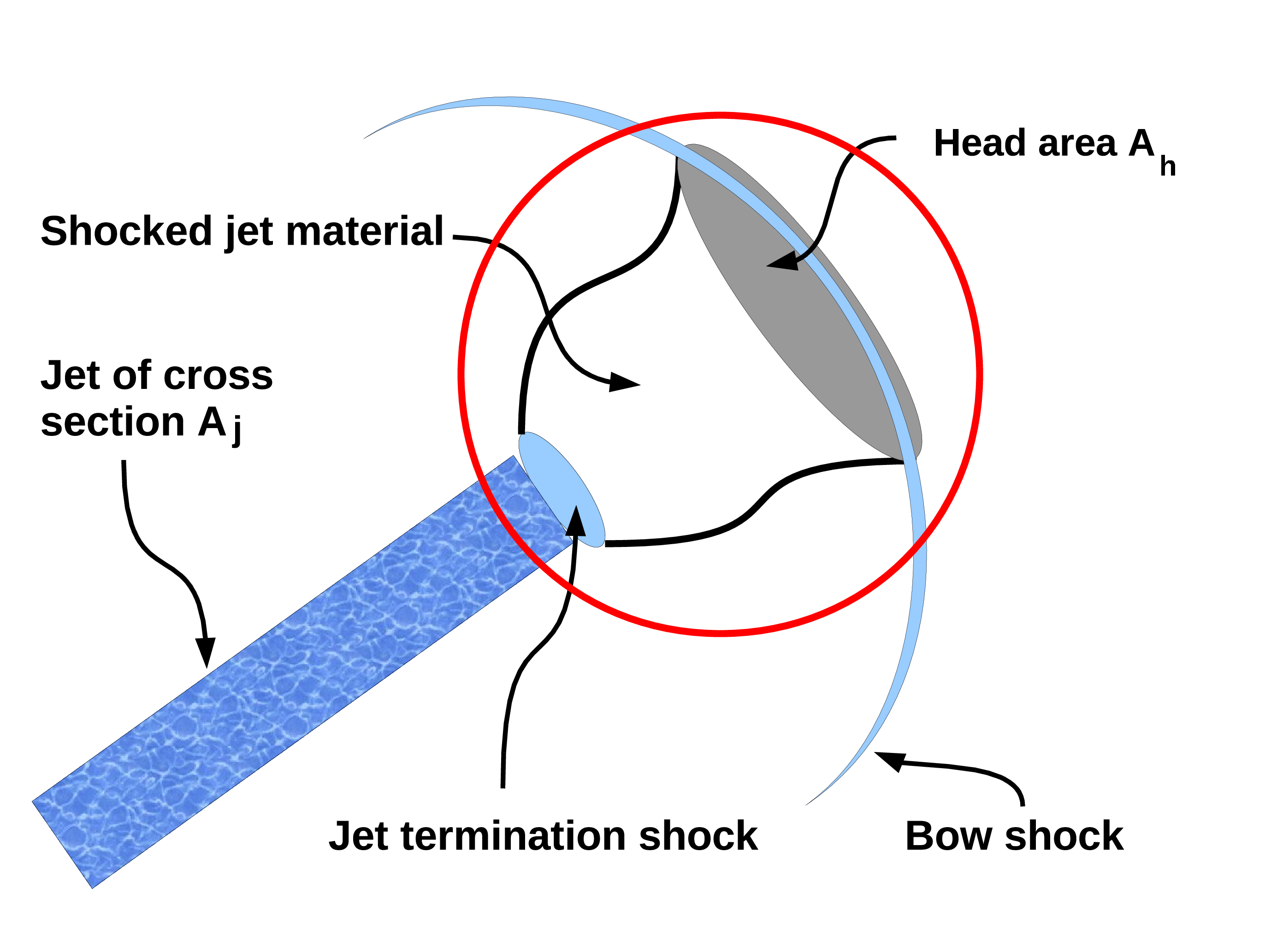,height=6.0cm,angle=0}
}
\caption[]{A schematic diagram of a hotspot-complex (encircled by the 
red cicle) at the head of an FR\,II jet.     
}
\label{fig_jet.shocks}
\end{figure}

Considering the conservation of momentum flow across the JTS and the
bow shock allows us to
find the hotspot velocity with respect to the host galaxy frame. 
Fig.~\ref{fig_jet.shocks}\footnote{\new{For simplicity we neglect the 
existence of multiple hotspots in this discussion, but that their 
existence implies that sometimes the deceleration (Gopal-Krishna \& Wiita 1990) 
and particle acceleration (Hardcastle, Croston \& Kraft 2007) may 
be distributed over a larger region than would be expected in a 
simple planar shock model. However, this makes no difference to 
our argument.}}
shows the schematic diagram of a jet head.   
Here we discuss the momentum balance at the hotspot, balancing the
momentum flux coming up the jet with the ram pressure at the front of
the lobe as it is driven through the ambient medium. The momentum balance equation can be written as 
(see equation~(\ref{eqn_mom.balance_2}) in Appendix)
\begin{equation}
\beta_{\rm hs} = \frac{1}{1+\eta}\beta_{\rm j},
\label{eqn_mom.balance}
\end{equation}
where $\beta_{\rm hs}$ and $\beta_{\rm j}$ are the hotspot velocity and jet bulk velocity 
in the host galaxy frame, and  
\begin{equation}
\eta = \sqrt{ \frac{\beta_{\rm j}cA_{\rm h}}{Q_{\rm j}}w_{\rm a}  }
\end{equation}
(from equation~\ref{eqn_eta}), where $A_{\rm h}$ is the area over
which the jet momentum flux is distributed, $c$ is the
speed of light, $Q_{\rm j}$ is the jet power as measured in the host
galaxy frame and $w_{\rm a}$ is the relativistic enthalpy density of
the ambient medium surrounding the jet and lobes.
The proper speed of the hotspot in the host galaxy frame, $u_{\rm hs}$, and
that of the jet matter in the hotspot frame, $u_{\rm j,hs}$ are relevant to
the discussion of the formation of the JTS and particle acceleration
there. These speeds are given by
\begin{equation}
u_{\rm hs} = \frac{\beta_{\rm hs}}{\sqrt{1-\beta^2_{\rm hs}}}
\label{eqn_prop.vel_hs}
\end{equation}
\begin{equation}
u_{\rm j,hs} = \Gamma_{\rm j,hs}\beta_{\rm j,hs} = \Gamma_{\rm j}\Gamma_{\rm hs}(\beta_{\rm j}-\beta_{\rm hs})  
\label{eqn_prop.vel_j_hs}
\end{equation}
In this formulation we can create the most general expression for the
ambient medium and, depending upon the situation, we can use the
values of the different parameters to distinguish between a thermal
and nonthermal ambient medium. In the outer lobes of DDRGs there can
be in general nonthermal particles ($e^{-}e^{+}$) together with a
(presumably small) amount of ingested thermal particles
($e^{-}p^{+}$). Similarly, in the thermal ambient medium of a radio
galaxy there can be in general thermal and nonthermal particles.
However, we know that the thermal particles in FR\,II radio lobes
(Croston et al. 2005) and the nonthermal particles in thermal ambient
medium of radio galaxies are not energetically dominant. We assume therefore
that the jet and lobe are composed of $e^{-}e^{+}$
plasma.

Therefore, the most general expression for $w_{\rm a}$, which contain both thermal and nonthermal 
components, can be written as 
\[w_{\rm a} = {\rm rest~mass~energy + remaining~internal~energy + pressure}  \]
\[w_{\rm a} = \left[n_pm_pc^2 + n_em_ec^2 \right] ~~~~~~~~~~~~~~~~~~~~~~~~~~~~~~~~~~~~~~~~~~~~~~~~~~~~~~~~~~~~~~~~~~~~~~~~~~~~~~~~~~  \]  
  \[+ \left[ \frac{3}{2}n^{th}_p k_BT^{th}_p+\frac{3}{2}n^{th}_ek_BT^{th}_e + \{(\epsilon^{nt}_e+\frac{B_{\rm a}^2}{8\pi})\}\right] ~~\]
\begin{equation}
     + \left[ n^{th}_pk_BT^{th}_p + n^{th}_ek_BT^{th}_e + \{\frac{1}{3}(\epsilon^{nt}_e + \frac{B_{\rm a}^2}{8\pi})\}\right] 
\label{eqn_enthal.in.gen}
\end{equation}
where $n_p=n^{th}_p$ is the number density of the thermal protrons only (and no nonthermal protons),
      $n_e=n^{th}_e+n^{nt}_e$ is the total number density of leptons (thermal + nonthermal), 
      $\epsilon^{th}=\frac{3}{2}n^{th}k_BT^{th}$ is 
      the kinetic energy density of the thermal particles,
      $P^{th}=n^{th}k_BT^{th}$ is the partial pressure 
      of the thermal partilcles,
      $\epsilon^{nt}_e$ is the energy density of the nonthermal electrons, 
      and $B_{\rm a}$ is the magnetic field strength in a general ambient medium. 

For nonthermal electrons which are ultrarelativistic, we define nonthermal electron temperatures (this is not a 
thermodynamic temperature) such that they follow
\begin{equation}
P^{nt}_e=n^{nt}_ek_BT^{nt}_e,
\label{eqn_pressure_nt_e}
\end{equation}
and 
\begin{equation}
\epsilon^{nt}_e=\frac{1}{\gamma -1}P^{nt}_e = 3n^{nt}_ek_BT^{nt}_e,
\label{eqn_int.energy_nt_e}
\end{equation}
where $\gamma=\frac{4}{3}$ is the adiabatic index of the gas
consisting of those relativistic particles.
Equations~\ref{eqn_pressure_nt_e} and \ref{eqn_int.energy_nt_e}
satisfy the relations $P=\frac{1}{3}\epsilon$. For thermal gas
$\gamma=\frac{5}{3}$, for a nonthermal distribution of
ultrarelativistic particles $\gamma=\frac{4}{3}$. If the nonthermal
particles are nonrelativistic or mildly relativistic then we cannot
characterise the gas by any constant adiabatic index. However, since
the particles are sufficiently scattered by the irregualrities of the
magnetic fields, we expect that such a gas is likely to have an
adiabatic index in the range $\frac{4}{3} < \gamma <
\frac{5}{3}$.

For the SDRGs and the outer doubles of the DDRGs the ambient medium is
the thermal gas around the radio galaxy. This means that $n^{nt}_p=n^{nt}_e=0$ (so
$\epsilon^{nt}_p=0$, $\epsilon^{nt}_e=0$) and $B_a=0$ are good
approximations. An expression for the enthalpy density, as obtained
from the equation~(\ref{eqn_enthal.in.gen}), is
\[w^{th}_{\rm a} = \left[n^{th}_pm_pc^2 + n^{th}_em_ec^2 + \frac{5}{2}n^{th}_p k_BT^{th}_p +  \frac{5}{2}n^{th}_e k_BT^{th}_e \right]  \]
\begin{equation}
      = \left[n^{th}_pm_pc^2 + n^{th}_em_ec^2 + \frac{5}{2}(n^{th}_p + n^{th}_e)k_BT^{th} \right]. 
\label{eqn_wa_th}
\end{equation}
Here we have substituted $T^{th}_p=T^{th}_e=T^{th}$, as electrons and
protons are usually in thermal equilibrium with each other in the
thermal environment of SDRGs.

For the inner doubles of double-double radio galaxies, the ambient
medium is the cocoon matter of the outer double and is dominated by
nonthermal matter. There might or might not be cold matter in the
outer cocoon of the DDRGs, but in the light of the observational
results of Croston et al. (2004, 2005), and as discussed above, there
is no evidence for significant thermal matter inside the radio lobes.
For the outer cocoon, we therefore assume that $n^{th}_p\sim0$ and
$n^{th}_e\sim0$. The expression for the enthalpy density, as
obtained from equation~(\ref{eqn_enthal.in.gen}), is then
\begin{equation}
w^{nt}_{\rm a} = n^{nt}_em_ec^2  +  \frac{4}{3}(\epsilon^{nt}_e +  \frac{B_{\rm a}^2}{8\pi}). 
\label{eqn_wa_nt}
\end{equation}

\subsection{Relativistic jet dynamics}
 It is the jet dynamics which control the hotspot speed, and
  therefore the strength of the JTS and bow shock. From our momentum
  balance analysis (Section~\ref{sec_mom.balance}), we know that the
  nature of the ambient medium is crucial in determining many aspects
  of jet dynamics. However, the ambient media for the inner jets and
  the outer jets of a DDRG are quite different. The outer jets and
  lobes propagate through a comparatively dense thermal medium,
  whereas, the inner jets and lobes propagate through a very tenuous
  non-thermal medium. We will discuss the effect of such differing
  ambient media on the jet dynamics in what follows. 

\subsubsection{Variation of hotspot speed with various parameters}
From equation~(\ref{eqn_mom.balance}), we know that the hotspot speed
depends on the parameter $\eta$ which itself is a function of various
other parameters, such as $\beta_{\rm j}$, $A_{\rm h}$, $w_{\rm a}$
and $Q_{\rm j}$. In this subsection we show graphically how the proper speeds of
hotspot motion measured in the host galaxy frame ($u_{\rm h}$) and the bulk jet
motion measured in the hotspot frame ($u_{\rm j,hs}$) change with the number
density of particles (protons) of the ambient medium, with
$\Gamma_{\rm j}$, $B_{\rm a}$, $r_{\rm h}$ and $Q_{\rm j}$ being treated
as parameters.

In Fig.~\ref{fig_betah.vs.np_plot_1}, $u_{\rm h}$ vs. $n^{th}_{\rm a}$
and $u_{\rm j,hs}$ vs. $n^{th}_{\rm a}$ curves are shown for two
different values of $B_{\rm a}$ in the left panel. A zoomed portion of
the left panel is shown in the right panel. For both panels we have
assumed a jet Lorentz factor $\Gamma_{\rm j}=5$ (as measured in the host glaxy
frame), a temperature of the thermal particles of the ambient medium
$T^{\rm th}_{\rm a}=3\times 10^{7}$ K, a jet head radius, $r_{\rm
  h}=5$ kpc, and a jet power, $Q_{\rm j}=10^{44}$ erg s$^{-1}$. The
thick continuous red curves are for the variation of $u_{\rm h}$ and
$u_{\rm j,hs}$ of the jets propagating through an ambient medium which
has neither magnetic field nor nonthermal particles; hereafter we
refer to this sort of ambient medium as a `pure thermal ambient
medium', and it is a good approximation for the medium surrounding the
outer double of a DDRG or any SDRG. The thin green dashed curves are
for the variation of $u_{\rm h}$ and $u_{\rm j,hs}$ of the jets
propagating through an ambient medium with thermal matter, constant
magnetic field $B_{\rm a}=$5 $\mu G$, nonthermal radiating electrons
in energy equipartition with the magnetic field (and no nonthermal
(jet-fed) protons). Hereafter, we refer to this sort of ambient medium
as an `impure ambient medium'. In the impure ambient medium the number
density of nonthermal electrons does not matter as their rest mass
energy is negligible compared to their kinetic energy and the rest
mass energy of the thermal protons. A third kind of ambient medium
contains only nonthermal particles (electrons and positrons) and
magnetic field, but no thermal matter; we can refer to this kind of
ambient medium as a `pure nonthermal ambient medium'. Though the thin
green curves are for an impure ambient medium, as we go towards very
low particle density, e.g. $n^{th}_p < 10^{-8}$ for the case in
Fig.~\ref{fig_betah.vs.np_plot_1}, the nonthermal energy (nonthermal
particle energy + magnetic field energy) starts dominating the thermal
energy, and for even lower values of $n^{th}$ an impure ambient medium
becomes equivalent to a pure nonthermal ambient medium, which is
perhaps the case for the inner doubles of DDRGs.
\begin{figure*}
\hbox{
 \psfig{file=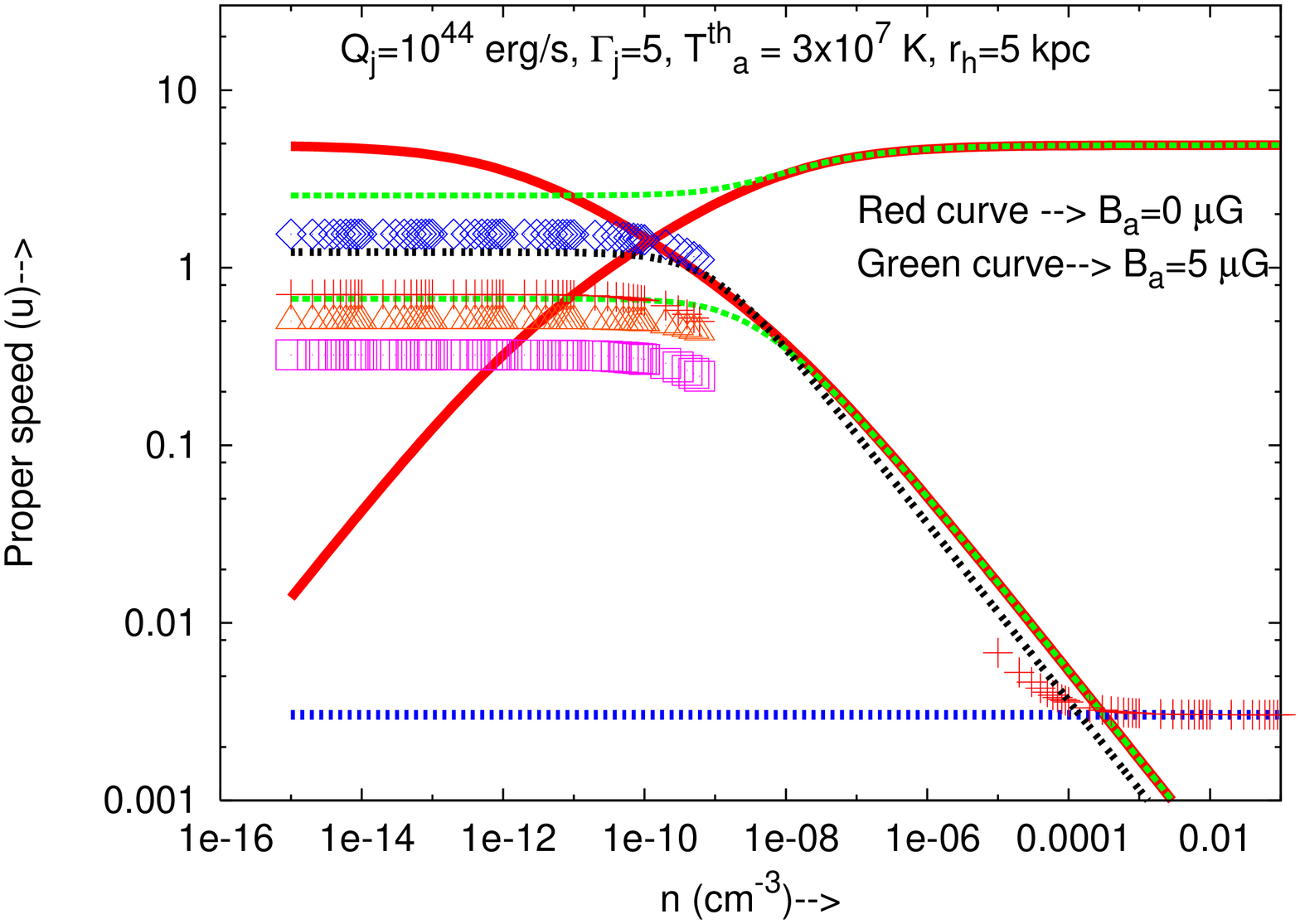,height=6.3cm,angle=-90}
 \psfig{file=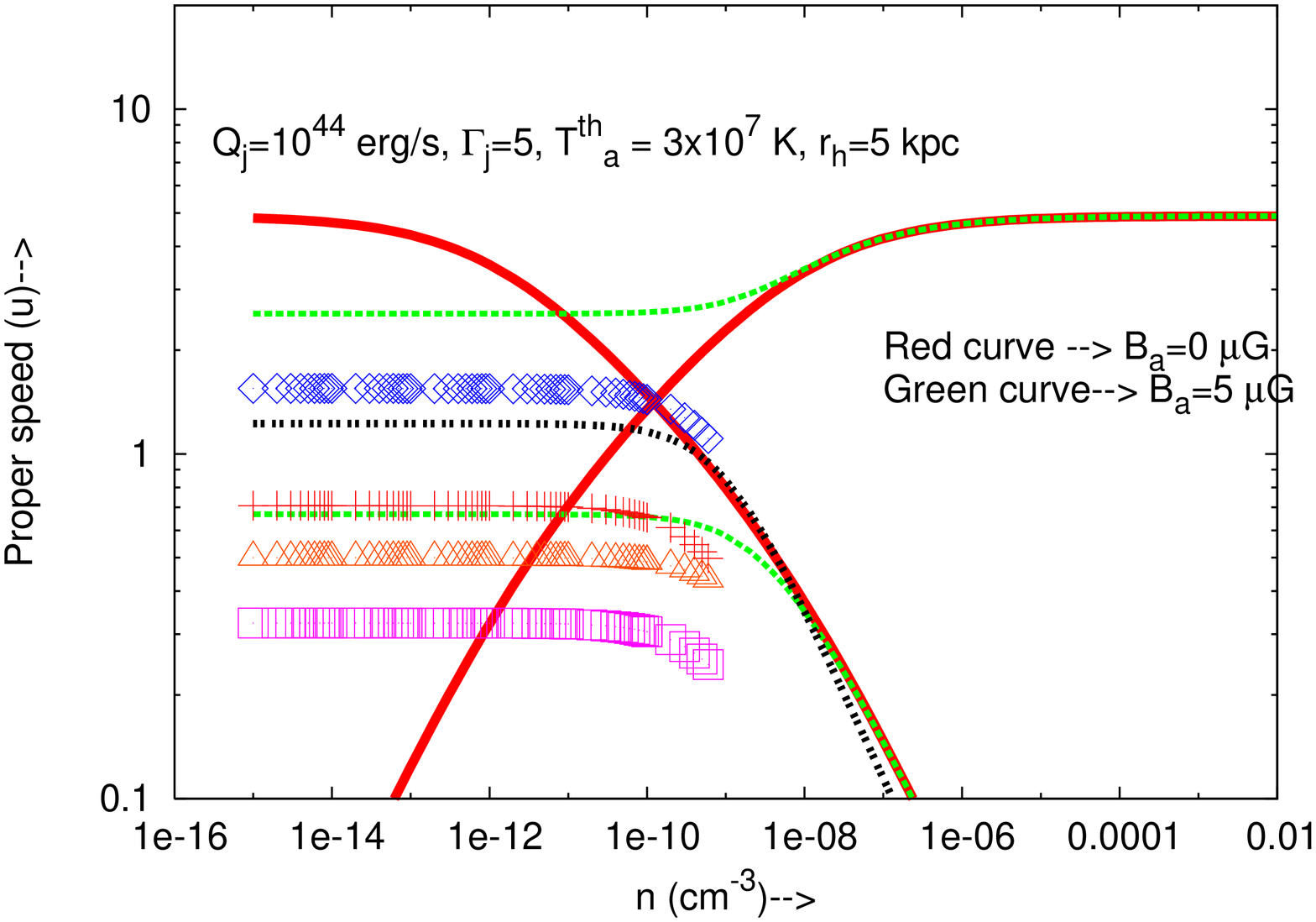,height=6.3cm,angle=-90}
}
\label{fig_betah.vs.np_plot_1}
\caption[]{ Left panel: the variation of proper speed of the hotspot
  motion ($u_{\rm h}$) as measured in the host galaxy frame and the upstream jet
  bulk motion as measured in the hotspot frame ($u_{\rm j,hs}=\Gamma_{\rm
    jh}\beta_{\rm jh}$) as a function of the number density ($n^{th}_p$) of
  thermal particles (protons) in the ambient medium. The thick red
  curves show the variation of $u_{\rm h}$ and $u_{\rm j,hs}$ with
  $n^{th}_p$ in a `pure thermal ambient medium'. The thin dashed green
  curves show the variations of $u_{\rm h}$ and $u_{\rm jh}$ for an
  `impure ambient medium' with constant magnetic field, $B_{\rm a}=$5
  $\mu G$ and no jet-fed protons. Among the red and green curves, the 
  ones which are rising as $n^{th}_p$ decreases are showing the variation 
  of $u_{\rm h}$, and the others are showing the variation of $u_{\rm
  j,hs}$. Equipartition between electron and field energy has been
  assumed. For all the curves a jet head radius $r_{\rm h}=5$ kpc,
  jet Lorentz factor $\Gamma_{\rm j}=5$, ambient medium thermal
  temperature $T_{\rm a}=3\times 10^7$ K, and jet power $Q_{\rm
    j}=10^{44}$ erg s$^{-1}$ have been used in the calculations. The
  black double-dotted curve shows the variation of the Alfven
  speed. The blue dotted horizontal line in
  the left panel is the sound speed in a `pure thermal ambient
  medium'. The other symbols indicates the variation of the
  magnetosonic wave mainly in the `pure nonthermal ambient medium'.
  Those are as follows. Blue diamonds: the speed of the fast
  magnetosonic wave; orange triangles: the speed of the
  intermediate magnetosonic wave (or shear Alfven wave); and pink
  squares: the slow magnetosonic wave. Red plus ($+$): the sound 
  speed in the absence of magnetic field. Note that we have drawn 
  the sound speed only in the ultrarelativistic and nonrelativistic 
  regime, as there is no simple method of calculating the appropriate
  adiabatic index in between these two regimes. 
  Right panel: a zoomed-in version
  of a portion of the same plot. }
\label{fig_betah.vs.np_plot_1}
\end{figure*}

\subsubsection{Propagation of jets through impure ambient media: the case of inner doubles}
\label{sec_inner.double}
\begin{figure*}
\hbox{
 \psfig{file=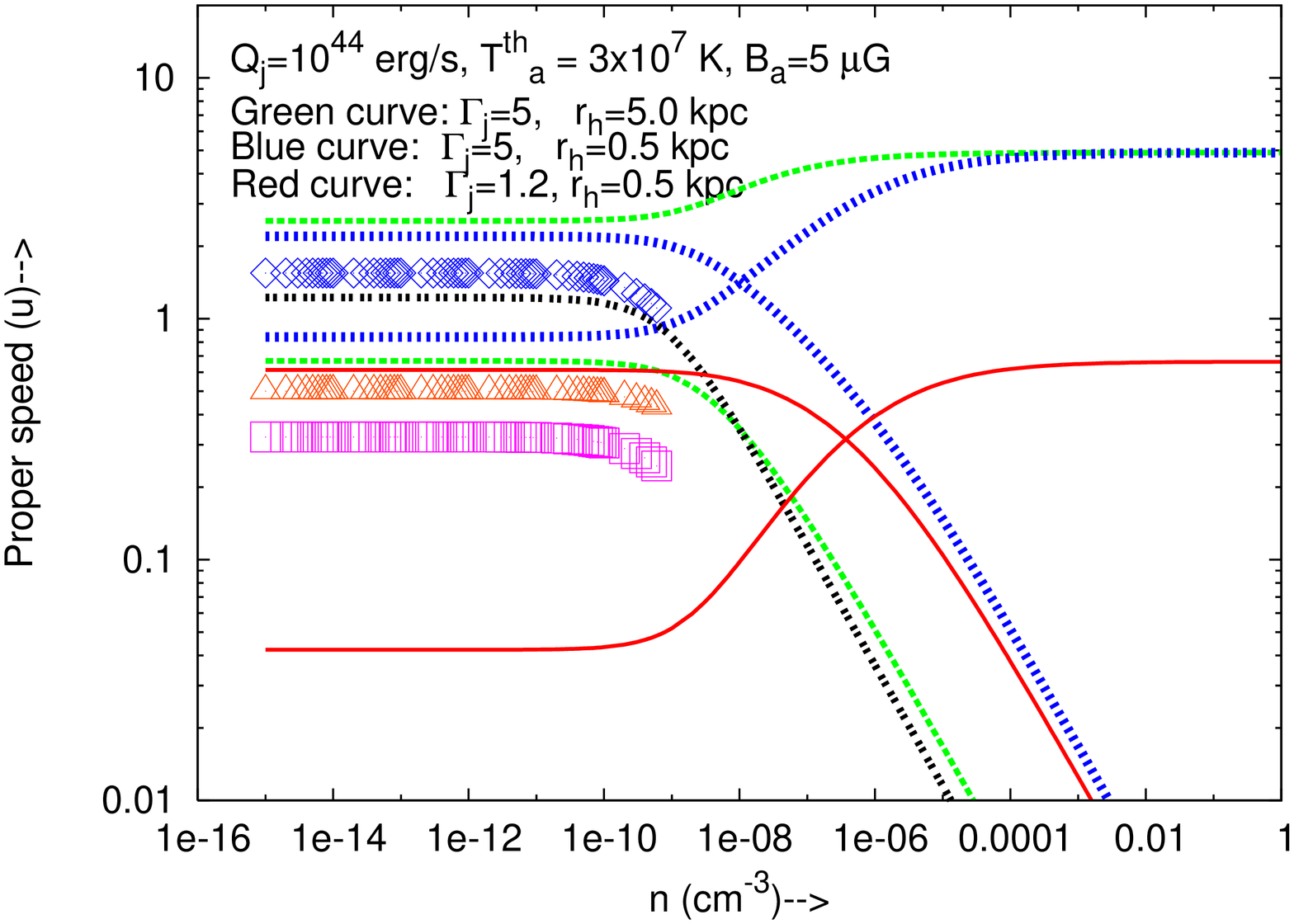,height=6.3cm,angle=-90}
 \psfig{file=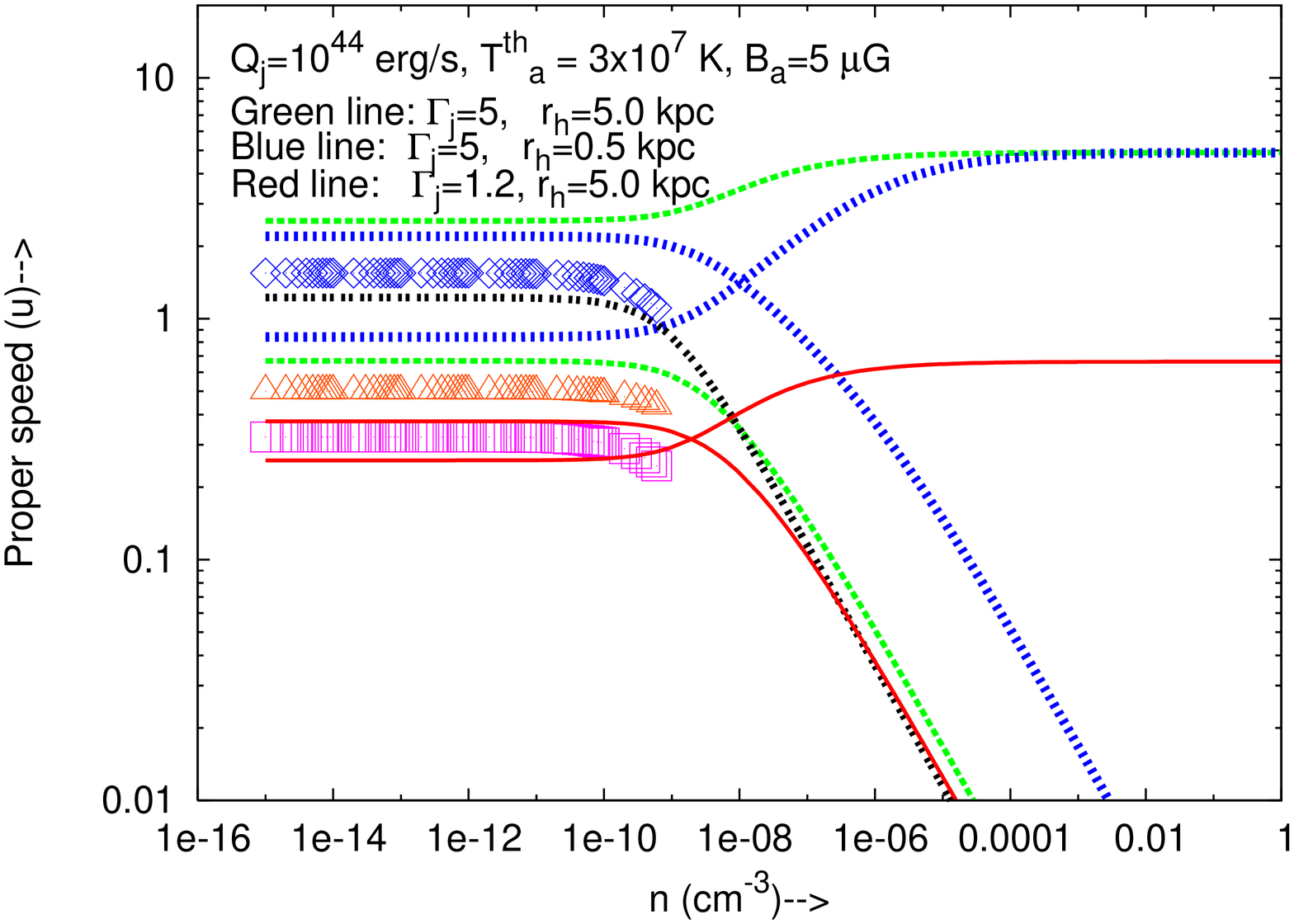,height=6.3cm,angle=-90}
}
\caption[]{ The variation of proper speed of the hotspot motion
  ($u_{\rm h}$) as measured in the host galaxy frame and the upstream jet bulk
  motion as measured in the hotspot frame ($u_{\rm j,hs}=\Gamma_{\rm
    j,hs}\beta_{\rm j,hs}$) as a function of the number density
  ($n^{th}_p$) of thermal particles in the ambient medium. All the
  curves are for $Q_{\rm j}=10^{44}$ erg $s^{-1}$, $T^{th}_{\rm
    a}=3\times 10^7$ K, $B_{\rm a}=5$ $\mu$G. The curves that fall
  and saturate as $n^{th}_p$ decreases show the variation of $u_{\rm
    j,hs}$ with $n^{\rm th}$. The curves that rise (except the
    double-dotted black curve) as $n^{th}_p$ decreases show the
  variation of $u_{\rm h}$. The green and blue curves are for
  $\Gamma_{\rm j}=5$ and the red curves are for $\Gamma_{\rm j}=1.2$. The
  rest of the symbols have the same meaning as in
  Fig.~\ref{fig_betah.vs.np_plot_1}. The green curves are for $r_{\rm
    h}=5$ kpc. The blue curves are for $r_{\rm h}=0.5$ kpc. The red
  curves on the left panel have $r_{\rm h}=0.5$ kpc, and on the right
  panel have $r_{\rm h}=5$ kpc. For simplicity, the variation of
    the speed of pure sound waves is 
  omitted in this plot.} 
\label{fig_betah.vs.np_plot_2}
\end{figure*}

In Fig.~\ref{fig_betah.vs.np_plot_2}, $u_{\rm h}$ vs. $n^{th}_p$ and
$u_{\rm jh}$ vs. $n^{th}_p$ curves are shown for two different values
of jet head radius, $r_{\rm h}$ in each panel. In this study we have
assumed an impure ambient medium with $B_{\rm a}=5$ $\mu$G,
$T^{th}_{\rm a}=3\times 10^7$ K. We have also fixed the jet power to
$Q_{\rm j}=10^{44}$ erg s$^{-1}$. The smaller the jet head radius, the
faster the jet head (or hotspot) can move through a given ambient
medium, provided all other parameters are held constant. This is quite
natural and conforms with our common sense that a thinner object can
penetrate easily through any given medium. The faster the speed of the
hotspot, the slower is the jet bulk speed, as observed from the hotspot
frame. Fig.~\ref{fig_betah.vs.np_plot_2} shows the variation $u_{\rm
  h}$ (and $u_{\rm jh}$) vs. $n^{th}_p$ curves for two different
values of $\Gamma_{\rm j}$. Keeping all other parameters fixed, if
$\Gamma_{\rm j}$ decreases from 5 to 1.2 then the entire curves
showing the variations of $u_{\rm jh}$ are displaced downwards. The
jet Lorentz factor thus has an important role in creating the JTS as well as
the bow shock around the jet head (see
Section~\ref{sec_JTS.n.bowshock} for detail).

\subsubsection{Various plasma waves in the nonthermal cocoon plasma}
\label{sec_plasma.waves}
In a magnetized plasma, disturbances can propagate via either pure sound waves 
(parallel to the field lines) and three kinds of magnetosonic waves. Those 
magnetosonic waves are fast magnetosonic wave (herefter, fast wave)
Intermediate wave (or shear Alfv\'en wave) and slow magnetosonic wave (hereafter, 
slow wave). For the general case of relativistic plasma with anisotropic pressure 
(the pressure parallel and perpendicular to field lines are different), the 
dispersion relations for magnetosonic waves have been worked out by 
Gedalin (1993). However, we know that the radio jets and lobes are
moving either 
through a nonrelativistic thermal plasma (for outer doubles of DDRG or SDRGs), 
or through an ultrarelativistic magnetized nonthermal plasma (for inner 
doubles of DDRGs). We have already referred to these as a `pure thermal ambient
medium' and a `pure nonthermal ambient medium' respectively. The adiabatic 
Equation of State (EoS) of both kinds of plasma can be written with a polytropic 
index, $\gamma_{\rm ad}$, as 
\begin{equation}
P \propto \rho^{\gamma_{\rm ad}},
\end{equation}     
where $\gamma_{\rm ad}=\frac{5}{3}$ for pure thermal ambient medium
and $\gamma_{\rm ad}=\frac{4}{3}$ for pure nonthermal ambient medium.
This polytropic EoS can be used for our purpose. A point to note here
is that even if the plasma is mildly to moderately relativistic, i.e.,
between nonrelativistic and ultrarelativistic limits, the adiabatic
EoS of the plasma can be still treated as a monatomic polytropic gas
with $\frac{4}{3} <\gamma_{\rm ad} < \frac{5}{3}$ (Synge 1957; Anile
1989). However, the problem is that in the case of a strong
relativistic shock, the adiabatic index in the downstream may be
different from that of the upstream flow. In our specific case,
though, we have enough
evidence, as already discussed, that the cocoon gas in outer doubles
of FR\,II DDRGs (or in general FR\,II lobes) can be treated as an
ultrarelativistic gas. Therefore an adiabatic EoS with a polytropic
index of $\frac{4}{3}$ can well describe the adiabatic behaviour of
the relativistic plasma in FR\,II lobes. Even if there are some
thermal/nonthermal protons in the radio lobes, we have shown in
equation~(\ref{eqn_Ep.lt.Ee}) that the kinetic energy in protons is
much less than that in the radiating electrons. This means that most of the
pressure is contributed by the radiating particles and magnetic field in FR\,II radio
lobes. Since the radiating particles are ultrarelativistic, an adiabatic EoS
with a polytropic index $\gamma_{\rm ad}=\frac{4}{3}$ is guaranteed.
If jet-fed protons and ingested thermal matter exist in cocoon matter,
they can only contribute to the mass density of the plasma. As far as
the lobe plasma is concerned, we know that observations show that
magnetic fields are not completely ordered (i.e. the degree of
polarization observed in lobes is well below the maximal possible
value) and so for simplicity, we will assume that the magnetic
field lines are completely tangled, which means that the
pressure of such a magnetoplasma is isotropic.

The EoS of the FR\,II jet matter is not known, but fortunately we do
not need to know it, as the EoS of upstream fluid is immaterial for a
strong relativistic shock. Since we know that JTS, at least in the
large-scale lobes, are associated with efficient localized particle
acceleration, it is reasonable to assume that all JTS are strong
shocks.

Chou \& Hau (2004) give a description of relativistic anisotropic MHD
for a gyrotropic ultrarelativistic plasma. Their work is directly
relevant for our case as they derived the dispersion relations for a
gyrotropic plasma rather than a general plasma, as given e.g. by
Gedalin (1993). The expressions for phase velocities of magnetosonic
waves, from Chou \& Hau (2004), reduce to the standard ones for a
non-relativistic, isotropic and polytropic plasma. Since the cocoon
plasma is very likely to exert isotropic pressure, we consider only
the isotropic case from Chou \& Hau (2004). Below we present the
expressions of the speeds of various waves in a `nonthermal cocoon
plasma' whose pressure is dominated by nonthermal radiating particles.

The Alfv\'en wave speed is given by
\begin{equation}
V^2_A = \frac{\frac{B^2}{4\pi}}{\rho^0c^2 + (\epsilon^{nt}+P^{nt}+\frac{B^2}{4\pi})}c^2
\label{eqn_sound.speed}
\end{equation}
and the pure sound wave speed in the absence of a magnetic field is given by 
\begin{equation}
V^2_{\rm son} = \frac{\gamma_{ad}P^{nt}}{\rho^0c^2 + (\epsilon^{nt}+P^{nt})}c^2
\label{eqn_Alfven.speed}
\end{equation}
(Chou \& Hau, 2004).  
The phase speed of the intermediate wave is given by (Chou \& Hau 2004)
\begin{equation}
V^2_I = \frac{\frac{B^2}{4\pi}\cos^2\theta}{\rho^0c^2 + (\epsilon^{nt} + P^{nt} + \frac{B^2}{4\pi})},
\label{eqn_Interm.speed}
\end{equation}
and the phase speed of the fast and slow wave mode can be given by 
\begin{equation} 
V^2_{\rm f,s} = \frac{1}{2}(b^{'} \pm \sqrt{{b^{'}}^2 - 4C^{'}}),
\label{eqn_magson.speed}
\end{equation}
where $b^{'}= b_{\rm is}/a_{\rm is}$ and $C^{'}= C_{\rm is}/a_{\rm
  is}$; these constants are described in Appendix~\ref{sec_A2}. For estimating numerical values for the plots 
in Fig.~\ref{fig_betah.vs.np_plot_1} and \ref{fig_betah.vs.np_plot_2}, we have used 
$\langle \cos^2\theta \rangle=\frac{1}{3}$.

We have used the
equations~(\ref{eqn_sound.speed})-(\ref{eqn_magson.speed}) to
calculate the speed of sound, Alfven and magnetosonic waves of the
realistic nonthermal plasma of the outer cocoon of a hypothetical DDRG
for our study and to generate the plots in
Fig.~\ref{fig_betah.vs.np_plot_1} and \ref{fig_betah.vs.np_plot_2}. On
the other hand, we have used equation~(\ref{eqn_prop.vel_hs}) and
(\ref{eqn_prop.vel_j_hs}) to estimate $u_{\rm hs}$ and $u_{\rm j,hs}$
respectively. Energy equipartition between radiating particles and
magnetic field has been used; inverse-Compton observations show that
this is a good approximation (Croston et al. 2005), as far as the
FR\,II lobes are concerned.

\subsubsection{Creation of jet termination shock and bow shock in the inner double}
\label{sec_JTS.n.bowshock}
Any FR\,II jet can, in principle, create two shocks. One is the JTS,
which is due to the sudden slow down of the jet flow when the jet flow
encounters the slow-moving downstream lobe material. The other is the
bow shock in the ambient medium around the front of the lobe. In other
works, the JTS
develops in the jet/lobe matter itself, whereas the bow shock develops
in the ambient medium through which the lobe ploughs its way. Since
both the jet matter as well as the nonthermal ambient medium of the
inner jets of a DDRG are relativistic magnetoplasma, we must compare
the speed of the bulk motions with the speeds of magnetosonic waves
rather than just sound waves. In any magnetoplasma a pure sound wave
without magnetic field disturbance can exist only for propagation
parallel to the field lines. The magnetic field energy density is
comparable to the kinetic energy density of the particles, hence the
magnetic field and its energy density cannot be neglected. Given that
the magnetic field is largely tangled, no pure sound wave without
magnetic field disturbance exists in the radio lobe plasma. In any
magnetised plasma there exist 3 magnetosonic waves as discussed in the
previous section. Therefore, there exist three types of shocks,
corresponding to bulk flows exceeding the speed of each of the
magnetosonic waves. The jet matter has a speed close to the
speed of light in the host galaxy frame. However, the strength of the JTS
depends on the speed of the jet matter with respect to the hotspot frame,
which is determined by the ram pressure balance equation given in
equation~(\ref{eqn_mom.balance}). As long as the ambient medium can
provide enough ram pressure so that the bulk speed of the upstream jet
matter in the hotspot frame is faster than at least the slow magnetosonic
wave, a JTS will form at the jet head in both inner and outer doubles
of any DDRG during their active phase. Our objective in this section
of the paper is to investigate whether this is indeed the case even
when the inner jets propagate through an extremely tenuous
outer-cocoon plasma containing no significant ingested thermal plasma.

We show the variation of the proper speeds of hotspots ($u_{\rm hs}$)
in the host galaxy frame and the upstream jet bulk motion ($u_{\rm j, hs}$) in the
hotspot frame in Fig.~(\ref{fig_betah.vs.np_plot_1}) and (\ref{fig_betah.vs.np_plot_2}), 
as usual assuming a jet power of $Q_{\rm j}=10^{44}$ erg s$^{-1}$. 
Let us consider the thick red curves in Fig.~(\ref{fig_betah.vs.np_plot_1}). 
These curves are for $u_{\rm j, hs}$ and $u_{\rm hs}$ (see the figure caption) 
propagating through a pure thermal medium with no magnetic field, as is 
the case for the outer double of a DDRG. A jet Lorentz factor $\Gamma_{\rm j}=5$ 
and a jet head radius $r_{\rm h}=5$ kpc have been used to derive the red curves. 
The sound speed of such a medium, given by
\begin{equation}
c_s = \sqrt{\frac{(5/3)k_BT^{th}}{\mu m_p}}
\end{equation}
is constant with density and is shown by a blue horizontal line in the
left panel of Fig.~\ref{fig_betah.vs.np_plot_1}. We have assumed a
poor cluster scale ambient medium of temperature $T^{th}=3\times10^7$
K. We can clearly see that $u_{\rm hs}$ is much faster than the proper
speed of sound in such an ambient medium. Therefore, the formation of
a bow shock is inevitable for this case. However, the formation of JTS
depends on the EoS of the jet material. For example, if the jet matter
is ultrarelativistic plasma then JTS formation is possible only for a
number density of protons in the ambient medium greater than $\sim
10^{-12}$ cm$^{-3}$ as evident from Fig.~\ref{fig_betah.vs.np_plot_1};
this condition is of course easily satisfied for all realistic
environments. Similarly, if the jet matter is pure thermal (which is
probably not the case) of temperature $T^{th}=3\times10^7$ K, then the
JTS will form up to a number density of protons of $\sim10^{-16}$
cm$^{-3}$ in the ambient medium.
 
When the same jet with same power and Lorentz factor propagates
through a pure nonthermal tenuous plasma medium (which is possibly the
case for the inner double of a DDRG) with cold protons (which
contribute to only mass density and not to pressure), then the
variation of $u_{\rm j, hs}$ and $u_{\rm hs}$ with the proton number
density $n_p$ of the ambient medium is shown by the green dashed
curves in Fig.~\ref{fig_betah.vs.np_plot_1}. We can see that for very
low densities ($n_p\lapp 10^{-8}$ cm$^{-3}$) of the ambient medium
both a JTS and a bow shock can easily form. In this case, whatever may
be the EoS of the jet material, the formation of JTS is inevitable, as
$u_{\rm j,hs}$ is faster than even the fast wave for a relativistic
EoS of the nonthermal plasma. We have studied this case for various
jet-head radii and Lorentz factors in
Fig.~\ref{fig_betah.vs.np_plot_2}. As is evident from that Figure, if
the jet matter has an ultrarelativistic EoS, then a jet with
$\Gamma_{\rm j}=1.2$ with both $r_{\rm h}=5$ and 0.5 kpc cannot form
JTS for $n_p\lapp10^{-10}$ cm$^{-3}$. The magnitude of this number is
of interest because the number densities of radiating electrons in the
outer lobes of FR\,II DDRGs are likely to be within the range
$10^{-9}-10^{-8}$ cm$^{-3}$. These plots show that, even without
protons, JTS and bow shocks can form (generally doing so more easily
for fast jets and large jet heads). However, if the lobe contains some
protons (through either entrainment or ingestion), then the JTS is
formed more easily. The key point from the results above is that
thermal matter entrainment/ingestion into the outer lobes is not
necessarily required to explain the confinement of inner lobes and the
formation of JTS in the inner jets of DDRGs, although we do not rule
out some amount of thermal matter ingestion into the lobes by some
means.

Our results are clearly different from those of Kaiser
  \etal\ (2000), who used a non-relativistic ram-pressure balance
  equation. Since the inner double expansion speed cannot be more
than $c$, Kaiser at al. (2000) had to restrict their ambient medium
density for the inner double. They cut off the curves (in their
Fig.~1), representing the variation of ambient medium density vs. jet
power below a certain density of the ambient medium. In their Fig.~1,
the continuous curves describing the amount of cocoon matter density
that is available from the $e^-e^+$ jet supply does not intersect with
the dashed curves describing the ambient medium density that can
reproduce the observed properties of the inner doubles for many DDRGs,
which has the effect that the jet is completely ballistic, if there is
only jet supplied matter in the cocoon. As a consequence, there is no
hotspot formation (and perhaps no confinement of the inner lobes as
well). However, this is an artefact of the application of the
nonrelativistic equation [see equation~(11) of Kaiser \& Alexander
  (1997)] for the momentum balance at the jet head. Whatever may be
the ambient medium density, the jet head can propagate at most at the
speed of the jet bulk motion as the ambient medium density tends to
zero. So ideally, in a correct model, the jet head speed, $\beta_{\rm
  h}$ should gradually approach $\beta_{\rm j}$ in the limit that
$\rho_{\rm a}\rightarrow 0$. This is definitely not the case in the
model of Kaiser et al. (2000). Because of their nonrelativistic
approach, the Kaiser et al. model artificially needs more cocoon
density to explain the observed hotspot formation and lobe confinement
of the inner double. In our model of momentum balance, it is quite
clear from equation~(\ref{eqn_mom.balance}) that when $\rho_{\rm
  a}\rightarrow 0$, $\eta \rightarrow 0$; therefore, $\beta_{\rm h}
\rightarrow \beta_{\rm j}$, hence our model has the correct behaviour
in the limit. Our fully relativistic equation~(\ref{eqn_mom.balance})
for momentum balance, for $\beta_{\rm h}\rightarrow 0$ limit for
denser $w_{\rm a}$, reduces to the widely used ram pressure balance
equation, $v_{\rm h}=\sqrt{\frac{Q_{\rm j}}{cA_{\rm h}\rho^0}}$.

We emphasise again that we do not rule out the possiblity of a small
amount of thermal matter entrainment/ingestion into the cocoons of
radio galaxies; low-frequency polarization observations are required
to search for it and the results can then be compared to those of
models such as those we present here.

\subsubsection{The hotspot motion of inner doubles is relativistic}
It is evident from the bottom panel of Fig.~\ref{fig_alpha-alpha_corrln}
that many of the DDRGs we have studied for our injection index related physics 
have outer jet powers greater than $10^{44}$ erg s$^{-1}$. As we will
argue in Section \ref{sec_discussion} the similarity 
of the injection indices (top panel of Fig.~\ref{fig_alpha-alpha_corrln}) 
in the outer and inner doubles suggests that the inner and outer jets
have similar power, so that it is reasonable to assume that our DDRGs
have {\it current} jet powers of order $10^{44}$ erg $s^{-1}$ or
higher. In this section we focus on DDRG with these jet powers, since
we have observational data to constrain our theoretical models. 
For the small sample of DDRGs that we have studied in our previous two papers 
(Konar et al. 2012, 2013) the typical magnetic field is $\gapp$5 $\mu$G
(except J0116-4722 which has $\sim0.27$ $\mu$G). This justifies the
assumption in Section \ref{sec_JTS.n.bowshock} of a jet power of
$10^{44}$ erg s$^{-1}$ and a magnetic field of 5 $\mu$G in the ambient 
medium.

\begin{figure*}
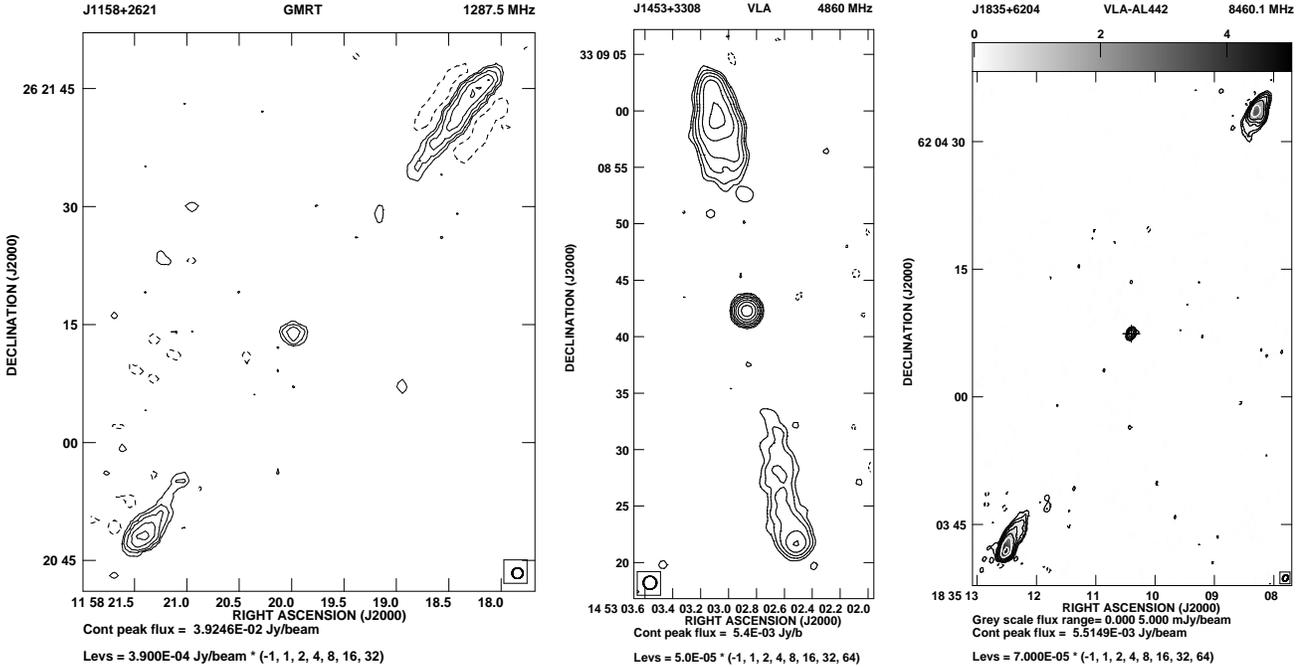

\hbox{
 \psfig{file=J1158L.INNER.PS,height=9.0cm,angle=0}
 \psfig{file=1450+333-INN-C.PS,height=9.0cm,angle=0}
 \psfig{file=J1835+6204_AL442.INN.DOUBLE.X.PS,height=9.0cm,angle=0}
}
\caption[]{The inner doubles of a few DDRGs.  The source name, telescope name and observed frequency
are given at the top of each image. The peak flux density, the countour level are given at the bottom
of each image. An ellipse within a square box at the bottom left or right corner is the resolution 
of the radio image. Left panel: This has been made with the GMRT data (project code: 10CKa01) with 
20 k$\lambda$ lower cutoff. This image has a resolution of $1.36\times1.36$ arcsec$^2$.  
Middle panel: This image is reproduced from Konar et al. (2006). Right panel: This image 
has been made with the VLA archival data (project code: AL442) and has a resolution of 
$0.79\times0.57$ arcsec$^2$ at a position angle of $\sim336^{\circ}$.   }
\label{fig_inner_double}
\end{figure*}

Our work in Section \ref{sec_JTS.n.bowshock} shows that the hotspots of the
inner jets have to travel at a the relativistic speed 
(see Fig.~(\ref{fig_betah.vs.np_plot_1}) and (\ref{fig_betah.vs.np_plot_2})), 
provided there is no significant entrainment/ingestion of thermal matter 
into the outer cocoon. We therefore predict that a relativistic 
beaming (or de-beaming) effect should be visible in the approaching (receding)
hotspots of inner doubles. The JTS in our radio images, with a
resolution of a few arcsec ($\sim 10$ kpc) will be seen as a point
source. It follows that any jet head emission 
can be thought of as two components: one is a point source component 
(the JTS), the other is the diffuse component which is the injected 
downstream plasma backflowing into the inner lobes. The beaming (or de-beaming)
effect should be prominently visible in the point source component. As
a result the inner jet-head which is approaching us should look like
a round standard hotspot and the inner jet-head which is receding from us
will have a shape more like the bow shock. Of course, to test this
prediction we need observations with an angular resolution which is enough to 
resolve our predicted structure, and yet does not resolve out the 
diffuse component with the given sensitivity of the telescope. We 
show high-resolution images of the inner doubles of a
few DDRGs in Fig.~\ref{fig_inner_double}, which, while obviously not
constituting a rigorous test of the model, show qualitative
agreement with our predictions, in the sense that one side of each
source has a more convincing compact hotspot than the other
(coincidentally, the S side in each of the three images shown here has
the more compact hotspot). We note that a pure relativistic model
would also predict that the approaching lobe would also be longer, due
to light-travel-time effects, which we do not observe here. In the
framework of the model, this would have to be explained in terms of
inhomogeneities in the outer cocoon material.

\begin{figure}
\hbox{
 \psfig{file=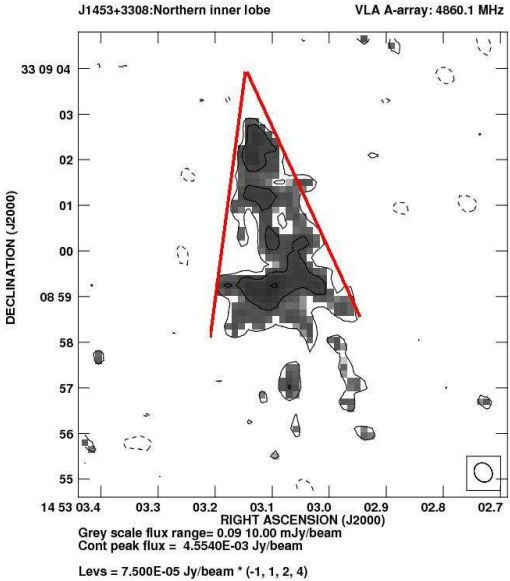,width=7cm,angle=0}
}
\caption[]{The inner northern lobe of J1453+3308. This is a C-band, A-array image with
resolution $0.44\times0.38$ arcsec$^{2}$ at Position angle of 36.03$^{\circ}$. The proposal
code of the data is AS697.}
\label{fig_N2-lobe}
\end{figure}

To investigate the Mach number in the inner doubles further, we have
made a high resolution image of the inner double of J1453+3308 from
the VLA archival data taken at C-band and with the A-array. As shown
in Fig. \ref{fig_N2-lobe}, the inner northern lobe shows a clear conical
structure. We estimate that the half angle of the cone is
$15.86^\circ$, which means that the relativistic Mach number would be
$M_{\rm rel} = 1/\sin(15.86) = 3.65$. This is consistent with the idea
that the northern lobe is a relativistic shock. Sophisticated
modelling, beyond the scope of the present paper, would be required to
find the possible thermal matter content within the outer lobes from
this estimate of $M_{\rm rel}$.

\section{Discussion}
\label{sec_discussion}
In the preceding section, we have addressed the theoretical aspects of the
inner jet dynamics of DDRGs. Keeping in mind the inner jet dynamics and the
theory of particle acceleration at the relativistic MHD shock, we discuss
the interpretation of our observational results in this section.

\subsection{Injection index and DDRGs}
\label{inj-index}
It is {\it a priori} quite surprising that the injection index is so 
similar between the two episodes of DDRG activity. Analytical modelling 
of particle acceleration at strong shocks suggests that the principal
factor determining the injection index should be the strength of the
JTS and the JTS strength depends on the proper speed of the upstream 
fluid ($\Gamma_{\rm j,hs}\beta_{\rm j,hs}$) as measured in the shock 
frame, i.e., hotspot frame (see Kirk et al. 2000). The relativistic 
velocity addition law yields
\begin{equation}
u_{\rm j,hs} = \Gamma_{\rm j,hs}\beta_{\rm j,hs}=\Gamma_{\rm j}\Gamma_{\rm hs}(\beta_{\rm j}-\beta_{\rm hs}),
\label{eqn_relvel}
\end{equation}
where $u_{\rm j,hs}$ is the proper velocity or the spatial components 
of the four velocity of the jet fluid as measured from the hotspot frame, 
$\Gamma_{\rm j,hs}$ is the Lorentz factor corresponding to the bulk speed, 
$\beta_{\rm j, hs}$ (in units of $c$) of the jet fluid as measured from the 
hotspot frame. From this equation, it is clear that the shock strength 
depends on $\beta_{\rm hs}$, the hotspot advance speed. Therefore, the JTS 
strength should depend on the ambient medium density ($\rho_{\rm a}$), 
as $\beta_{\rm hs}$ depends on $\rho_{\rm a}$
through ram pressure balance at the contact discontinuity at the head
of the lobe. However, if the ambient medium consists of relativistic
particles, as it does for the inner jets of the DDRGs, where the
ambient medium is the cocoon material of the outer doubles, then
$\rho_{\rm a}=\frac{\varepsilon}{c^2}$, where $\varepsilon$ is the
relativistic internal energy density which contains the rest mass
energy also, and $c$ is the speed of light. This should be at least 
two orders of magnitude less (Brocksopp et al. 2007, 2011; Safouris et
al., 2008; Clarke \& Burns, 1991) than that of the thermal ambient
medium through which the outer jets had propagated. $\beta_{\rm hs}$ 
should accordingly be faster in the inner hotspots, and the JTS
is expected to be weaker than that in the previous episode, assuming
constant $\beta_{\rm j}$ and $Q_{\rm j}$, an assumption that we revisit 
below. We might therefore naively expect to see systematic differences 
between the injection indices in the two lobes in DDRG, in the sense that 
$\alpha_{\rm inj}$ in the inner doubles compared to the outer doubles, 
which is not what we observe (except possibly in the case of 3C293: see
Fig.~\ref{fig_alpha-alpha_corrln})

What can account for this discrepancy between observation and
expectation? Since the strength of the JTS depends on the proper speed
(or the spatial components of the four-speed, $u_{\rm
  j,hs}=\Gamma_{\rm j,hs}\beta_{\rm j,hs}$) of the upstream jet flow
as measured in the hotspot frame, it is worth estimating the proper
velocities, $u_{\rm j,hs}$, of the upstream jet flow for a fixed
$\Gamma_{\rm j}\sim2$ (see Mullin \& Hardcastle 2009) and a range of
hotspot velocities, $\beta_{\rm hs}$. For the classical double radio
galaxies (which are equivalent for the outer double of DDRGs) the
hotspot velocities range from 0.05$-0.1c$ (Konar et al. 2009; Jamrozy
et al. 2008; Scheuer 1995) and for the inner doubles the hotspot
velocities range from 0.05$-0.5c$. The lower limit is not a very
stringent limit which is suggested by our data, published by Konar et
al. (2012; 2013), on the lower limits on the break frequency in the
synchrotron spectra of inner doubles. Since most of the sources do not
have deep images at frequencies greater than 8.4 GHz, we cannot
directly measure the break frequencies or even place a very stringent
limits in break frequency. The upper limit on the inner hotspot speed
is from Safouris et al., (2008). Assuming $\Gamma_{\rm j} \sim 2$,
which corresponds to $\beta_{\rm j} \sim 0.866$, the application of
equation~(\ref{eqn_relvel}) yields values of $u_{\rm j,hs}$ of 1.634,
1.539 and 0.665 for $\beta_{\rm hs}$ values of 0.05, 0.1 and 0.5
respectively. For the outer doubles, with their low hotspot advance
speeds, we find $1.539 \ga u_{\rm j,hs} \ga 1.634$, i.e. a very
similar proper velocity; for speeds as high as the upper limit on
$\beta_{\rm hs}$ for the inner doubles, the proper velocity is much
lower and so the JTS would be expected to be weaker. We conclude that
if the injection index$-$jet power correlation observed in classical
FR\,IIs were purely due to JTS strength, then the injection indices of
the inner doubles should have been steeper in all DDRGs, as inner
hotspots are very likely to move faster than their outer counterparts.
However, this is not what we observe.

There are two possibilities to explain these observations: either (i)
the jet bulk Lorentz factor has such a high value that the difference
between inner and outer double in $\beta_{\rm hs}$ does not cause
significant difference in shock strength and injection index, or (ii)
standard DSA models in the relativistic regime do not actually
describe the hotspots. We prefer the former possibility. Our study of
momentum balance at the jet head in Section~\ref{sec_JTS.n.bowshock}
shows that the formation of JTSs and bow shocks in inner jets is quite
plausible even if the ambient medium is a tenuous nonthermal medium
with no external thermal matter ingested into it. Of course, $u_{\rm
  j,hs}$ will be different for inner and outer jets during their
active phase; however, if the jet Lorentz factor is high enough then
the JTSs in both outer and inner jets will produce the universal
(true) injection index of 0.62 (Kirk et al. 2000), which will then be
steepened (by synchrotron losses) by the same amount when the plasma
comes out of the hotspots (to produce the same observed injection
index) provided the jet powers are the same in both episodes. The
higher the jet Lorentz factor, easier will be to explain the
$\alpha^{\rm inn}_{\rm inj}-\alpha^{\rm out}_{\rm inj}$ correlation.
We explore this type of model in the following subsections.

\subsection{Jet power and injection index}
\label{sec_jp-index}
Fig.~\ref{fig_alpha-alpha_corrln} shows that there is a strong
correlation between jet power and injection index in radio galaxies in
general, including some of the outer lobes of our DDRG sample. This
observation prompts two questions: (i) can this correlation account
for the correlation between the injection indices in the inner and
outer doubles in the DDRG sample, and (ii) how do these correlations
arise?

First of all, it is worth considering what determines the jet power of
a radio galaxy and whether any of the parameters in question are
expected to change between different episodes of activity. In general,
we expect the jet power to depend on properties of the accretion
system such as the black hole mass $M_{\bullet}$, the accretion rate
${\dot M}$ and perhaps the black hole spin $S_{\bullet}$ (e.g.
Blandford \& Znajek 1977; Bhattacharya, Ghosh \& Mukhopadhyay 2010).
Other parameters set by the jet generation mechanism, such as the
ratio ($\sigma_j$) of magnetic enthalpy density
($w_B=\frac{B^2}{4\pi}$) to particle enthalpy density (Kirk et al.
2000) may also play a role. 

Can the black hole of a DDRG increase significantly during the 
quiescent phase of jet forming activity? 
If we assume that during the quiescent phase of jet forming activity, 
the AGN accretes and radiates efficiently, then we can assume the 
canonical values of $\lambda=0.1$ and $\xi=0.1$. Radio galaxies are 
supposed to be produced by black holes of mass $\gapp10^8M_{\odot}$ 
(Chiaberge \& Marconi 2011). For a SMBH of $M_{\bullet}=10^8 M_{\odot}$ 
and the time scale of quiescent phase of jet forming activity of $10^7$ yr 
(or less for our sample of sources), the SMBH will increase its mass by 
$\Delta M = 2.27\times 10^6 M_{\odot}$ (using equation~\ref{eqn_tot.mass.accrn_2} 
in Appendix~\ref{sec_A3}). This is much smaller than the initial black hole 
mass which is $10^8 M_{\odot}$. This calculation is done for the higher end 
of the range of the quiescent phase and with the assumption that the AGN 
accretes efficiently via thin-disk accretion mode during this phase, and so 
is likely to be an overestimate for a typical DDRG. Therefore, it is clear 
that mass accretion cannot significantly change $M_{\bullet}$ of the central 
SMBH during the time scale of the quiescent phase which ranges from $\sim10^5$ yr
to a several tens of Myr (Konar et al. 2013; 2012; 2006; Saikia et
al., 2006).

Since the DDRGs we are concerned are all powerful FR\,II sources and are 
high excitation radio galaxies, they are thought to be accreting via a 
standard accretion disk (e.g. Hardcastle, Evans \& Croston 2006). 
Matter constantly loses angular momentum via the disk to fall into 
the black hole. However, inside the Innermost Stable Circular Orbit (ISCO), 
there is no stable orbit. The matter should thus fall from the ISCO to 
the SMBH with the angular momentum corresponding to the Keplerian motion 
at the ISCO. This assumption should give us an upper limit on the angular 
momentum deposited into the SMBH during the quiescent phase. We estimate 
the maximum angular momentum deposited in the following way. We assume
that the ISCO is at $3R_s$, where $R_s=\frac{2GM_{\bullet}}{c^2}$ is
the Schwarzschild radius. If the black hole spins, the ISCO is
supposed to be smaller, however our estimation of angular momentum
with $3R_s$ will give an upper limit, which we are interested in.
As long as the quiescent phase of jet forming activity is $\lapp 10^7$ yr, we do not
expect that the SMBH at the central engine will accumulate any
substantial amount of amgular momentum which will change the jet power
significantly (see Appendix~\ref{sec_A3}). This result will hold for any 
SMBH with any mass higher than $10^8 M_{\odot}$ for the same values of 
the other parameters. Therefore, it is likely that $M_{\bullet}$ and 
$S_{\bullet}$ do not change dramatically in two episodes of jet forming 
activity, unless there is a merger with another galaxy (Schoenmakers et al.\ 2000a), 
or SMBHs in a comparable-mass binary coalesce between two episodes (Liu et al.
2003). However, the stability of the jet axis (Konar et al. 2006;
Saikia et al.\ 2006) between the two episodes and the duration of the
quiescent phase of the DDRGs (Konar et al. 2012, 2013) are not at all
favourable for a merger with another galaxy (merger time scale
$\sim10^9$ yr). SMBH coalescence in unequal mass binary is quite
plausible between the two episodes, provided that the merging black
hole is small enough to maintain the stability of the jet axis between
the two episodes, but, as a consequence of that restriction, this also
does not affect $M_{\bullet}$ and $S_{\bullet}$. Finally, we note that
if the quiescent phase is due to lack of fuel (i.e. a large drop in
$\dot M$), it is clear that $M_{\bullet}$ and $S_{\bullet}$ must
remain unchanged in the two episodes of jet forming activity.

We have little direct information on $\sigma_j$, but for the
kpc/Mpc-scale sources the lobes and hotspots are close to
equipartition (Croston et al. 2005; Hardcastle et al. 2004), i.e.
$\sigma \sim 1$. It is plausible that the jet fluid is also in
equipartition. This means $\sigma_{\rm j}$ should be $\sim 1$ in both
the episodes.

The factor that can most easily change between the two episodes is the
fuel supply, $\dot M$. If the black hole is accreting from the hot-gas
environment, the accretion rate is expected to be determined by the
central cooling time, which is $\sim 10^8$ yr, so that the density of
hot gas is unlikely to change during the quiescent phase that we have
determined for these FR\,II DDRGs (though we note that Jetha et al.
(2009) report on a low-power system thought to be powered by hot-gas
accretion where the quiescence timescale is indeed $\sim 10^8$ yr).
But in a scenario where the fuel supply comes from cold gas, the
timescale for significant variations can be comparable to the
Keplerian timescale for the torus, which could be $<10^6$ yr. There
thus seems to be no physical reason why $Q_{\rm j}$ should be
correlated between different epochs of jet forming activity. However, 
in fact, there is
a very strong selection effect requiring the two values of $Q_{\rm j}$
to be at least comparable in the objects we study: if they are very
different, we will not observe a classical DDRG in which the inner and
outer doubles are of comparable luminosity. Thus, if we can explain
why $Q_{\rm j}$ controls the injection index, we can explain the
observed correlation between the injection indices of the two epochs
in terms of a correlation between their values of $Q_{\rm j}$.
\begin{table*}
\caption{Source parameters that are employed in plotting $\alpha_{\rm inj}-Q_{\rm j}$ 
(in Fig.~\ref{fig_alpha-alpha_corrln}) and $\alpha_{\rm inj}-z$ 
correlations respectively. The component designations are N-lobe (S-lobe) is the
northern (southern) lobe, NW-lobe (SE-lobe) is the North-western (south-eastern) lobe,
Int-core is the core subtracted integrated source.
The description of the columns is as follows. Column 1: source name,
Column 2: Alternative name, Column 3: redshift of the host galaxy, Column 4: injection
spectral index of the radio galaxies, Column 5: jet power and  Column 6: reference and comment.}
\label{tab_su.z.alpha.jetpow}
\begin{tabular}{l l l l l l r c r r c}
\hline
Source         &Alt.         &   z     & $\alpha_{\rm inj}$         & $Q_{\rm j}$          &  Ref.       \\
               &name         &         &                        & ($10^{45}$ erg s$^{-1}$) &   and       \\
               &             &         &                        &                          &  comment    \\
(1)            &  (2)        &  (3)    & (4)                    &  (5)                     &   (6)       \\
\hline
 DDRGs$^{\dag\dag}$     &                        &                & $\alpha^{out}_{\rm inj}$       & $Q^{out}_j$                &         \\ \hline
 J0041+3224             &  B2\,0039+32           & 0.45$^{\dag}$  & 0.756                          &   0.9315                   &    1    \\
 J0116-4722             &  PKS\,0114-47          & 0.14610        & 0.618                          &   $\ast$                   &    2    \\
 J0840+2949             &  4C\,29.30             & 0.064715       & 0.810                          &   $\ast$                   &    3    \\
 J1158+2621             &  4C\,+26.35            & 0.112075       & 0.788                          &   0.1656                   &    2    \\
 J1352+3126             &   3C293                & 0.045034       & 0.855                          &   $\ast$                   &    4    \\
 J1453+3308             &  4C\,33.33             & 0.248174       & 0.568                          &   0.3156                   &    5    \\
 J1548-3216             &  PKS\,1545-321         & 0.1082         & 0.567                          &   0.1934                   &    6    \\
 J1835+6204             &  B1834+620             & 0.5194         & 0.818                          &   2.8520                   &    1    \\ \hline
 Large radio galaxies    &                        &                & $\alpha_{\rm inj}$             &  $Q_j$                     &         \\ \hline
 J0135+3754 Int-core    & 3C46                   & 0.4373         & 0.982                          &  13.71                     &    7    \\
 J0912+3510 N-lobe      &                        & 0.2489         & 0.560                          &   0.12                     &    8    \\
 J0912+3510 S-lobe      &                        & 0.2489         & 0.628                          &   0.33                     &    8    \\
 J0927+3510 NW-lobe     &                        & 0.55$^{\dag}$  & 0.750                          &   2.00                     &    8    \\
 J0927+3510 SE-lobe     &                        & 0.55$^{\dag}$  & 0.700                          &   2.36                     &    8    \\
 J1155+4029 NE-lobe     &                        & 0.53$^{\dag}$  & 0.876                          &   5.94                     &    8    \\
 J1155+4029 SW-lobe     &                        & 0.53$^{\dag}$  & 0.838                          &   1.42                     &    8    \\
 J1313+6937 NW-lobe     &                        & 0.106          & 0.610                          &   0.16                     &    8    \\
 J1313+6937 SE-lobe     &                        & 0.106          & 0.610                          &   0.15                     &    8    \\
 J1343+3758 NE-lobe     &                        & 0.2267         & 0.570                          &   0.25                     &    8    \\
 J1343+3758 SW-lobe     &                        & 0.2267         & 0.570                          &   0.41                     &    8    \\
 J1604+3438 W-lobe      &                        & 0.2817         & 0.554                          &   0.13                     &    8    \\
 J1604+3438 E-lobe      &                        & 0.2817         & 0.554                          &   0.13                     &    8    \\
 J1604+3731 NW-lobe     &                        & 0.814          & 0.765                          &   2.71                     &    8    \\
 J1604+3731 SE-lobe     &                        & 0.814          & 0.775                          &   3.39                     &    8    \\
 J1702+4217 NE-lobe     &                        & 0.476          & 0.588                          &   0.64                     &    8    \\
 J1702+4217 SW-lobe     &                        & 0.476          & 0.588                          &   0.59                     &    8    \\
 J2245+3941 Int-core    & 3C452                  & 0.0811         & 0.782                          &   1.03                     &    7    \\
 J2312+1845 NE-lobe     & 3C457                  & 0.427          & 0.820                          &   5.59                     &    8    \\
 J2312+1845 SW-lobe     &                        & 0.427          & 0.820                          &   2.37                     &    8    \\
\hline
\end{tabular}
\begin{flushleft}
$\ast$: No definite age estimate exists, only a limit, so the source
  is not plotted. \\
$\dag$: Estimated redshift. \\
$^{\dag\dag}$: For DDRGs, we have considered the outer double for the $\alpha_{\rm inj}-Q_{\rm j}$ and $\alpha_{\rm inj}-z$ correlation. \\
1: Konar et al. (2012), 2: Konar et al. (2013), 3: Jamrozy et al. (2007), 4: Joshi et al. (2011), 5: Konar et al. (2006),
6: Machalksi et al. (2010), 7: Nandi et al. (2010), 8: Jamrozy et al. (2008).
\end{flushleft}
\end{table*}

We discuss in this section the mechanism by which jet power can affect
the injection index. First of all, as noted above, DSA models predict
that the original power law index will depend on the shock strength. A
higher jet power implies a higher jet momentum flux, which will drive
a faster expansion of the lobe (higher $\beta_{\rm hs}$) for a given
external density, and so would actually weaken the jet termination
shock, giving injection index. This weakening of JTS by higher
$Q_{\rm j}$ will work while $\Gamma_{\rm j}$ is not very high, because
for very high $\Gamma_{\rm j}$ the $\beta_{\rm j,hs}$ will always be
very close to 1, no matter what the value of $\beta_{\rm hs}$ is. So,
in the high-$\Gamma_{\rm j}$ regime, if a higher jet power implies a
higher jet bulk {\it speed} ($\beta_{\rm j}$) then the JTS will be
stronger and higher jet power will imply flatter injection index.
There are some observational constraints on this possibility: Mullin
\& Hardcastle (2009) recently found no statistical correlation between
radio power and $\Gamma_{\rm j}$ for a complete sample of 98 FR\,II
sources with $z<1$ spread over a monochromatic luminosity range of
$\sim3$ orders of magnitude at 178 MHz. Taken at face value, these
results imply that the jet bulk speed for large source sizes (as
studied by Mullin \& Hardcastle) is independent of the jet power, and
the $\Gamma_{\rm j}$ ($\sim2$) is not very high. This in turn implies,
by the argument of Section \ref{inj-index}, that the shock strength
might {\it decrease} for higher jet power, which is in the sense
required by the observations. This can explain, at least
qualitatively, the jet power$-$injection index correlation, relying on
the assumptions that (i) the jet speed across the cross section is the
same and close to the directly estimated value of
$\Gamma_{\rm j}$ (e.g., $\sim2$) and (ii) the external density is the
same for all sources (though this is probably not the case). However,
it certainly fails to explain the similar injection index in two
episodes for the following reasons. The dynamical modelling of Inner
doubles of J1548-3216 by Safouris et al. (2008) and the conical shape
(which is reminiscent of a bow-shock) of the inner northern lobe head
of J1453+3308 (see Fig.~\ref{fig_inner_double}) as shown by Konar et
al (2006) suggest much faster hotspot velocity and tenuous ambient
medium density compared to the outer doubles (yet the injection index
of the inner doubles are similar to that of the outer doubles). This
can only be possible if the principal parameters that injection index
depend on, e.g., $Q_{\rm j}$, $\sigma_{\rm j}$ and $\rho_{\rm a}$,
have adjusted values so as to cause no change in injection index in
the two episodes. Clearly, this is highly unlikely. Therefore, lower
value of $\Gamma_{\rm j}$ cannot explain the observations at all.

\begin{figure}
\vbox{
 \psfig{file=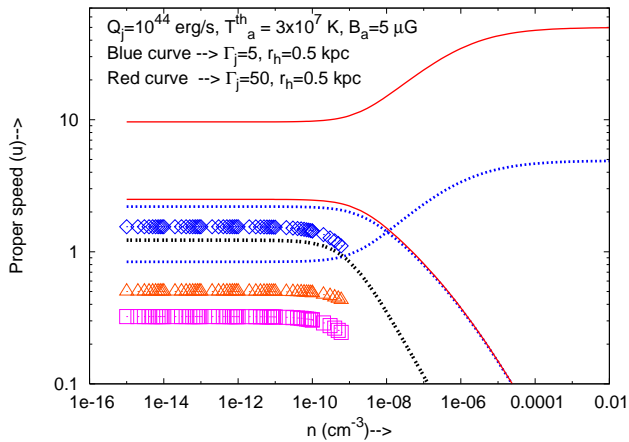,height=6.0cm,angle=-90}
}
\caption[]{Plots as in
  Fig.~\ref{fig_betah.vs.np_plot_2} for high bulk Lorentz factor jets.
  The red curves are for $\Gamma_{\rm j}=50$. See the caption of 
  Fig.~\ref{fig_betah.vs.np_plot_2} for the colours, symbols and 
  descriptions. The parameter values are annotated in the plot. }
\label{fig_betah.vs.np_plot_3}
\end{figure}

If the model described by Kirk et al. (2000) describes the particle
acceleration phenomena at the hotspots, then a possible curve of the
kind shown in Fig.~4 of their paper (the top panel of
Fig.~\ref{fig_betah.vs.np_plot_3} of this paper), but with an
appropriate value of $\sigma_{\rm j}$ (i.e., $\sigma_{\rm j}$ at
equipartition) with a very high Lorentz factor flow of jet fluid,
combined with higher synchrotron losses due to higher magnetic field at
the hotspots of high power sources, can explain the injection
index$-$jet power correlation as well as the the similarity of
injection index in two different episodes. However, keeping in mind
the low Lorentz factor jets suggested by Mullin \& Hardcastle (2009),
we hypothesize a structured jet with a fast moving spine inside the
jets, the bulk of the jet kinetic power is concentrated in
the spine and thus it dominates the dynamics and the particle
acceleration\footnote{In a completely independent argument,
    Hardcastle (2006) argues that to
  model the jet related X-ray emission of core-dominated radio loud quasars
  as inverse-Compton scattering of Cosmic Microwave Background (CMB)
  photons, the jet has to have velocity
  structure and the spine of the jet should have $\Gamma_{\rm j} \gapp
  15$. This is fully consistent with our results.}. To be consistent with 
Fig.~4 of Kirk et al. (2000), the spine can
be moving with very high Lorentz factor ($\ga 10$) and outer layers
denser than the spine of the jets move slower so as to produce a lower
Lorentz factor (e.g. $\sim2$) as estimated by Mullin \& Hardcastle
(2009). Since the particles from the high Lorentz factor spine are
accelerated at the JTS, the ambient medium density does not affect the
value of injection index provided that the jet power (with jet fluid being
in equipartition) remains same in both episodes. We have plotted the
curves showing the variation of $u_{\rm j,hs}$ (and $u_{\rm h}$) vs.
$n_{\rm a}$, i.e., the number density of particles in the ambient
medium for two different values of the Lorentz factor (5 and 50)
keeping the same jet head radius of 0.5 kpc and the values of the
other parameters. Since from Kirk et al. (2000) (see top panel of
Fig.~\ref{fig_betah.vs.np_plot_3}), it is clear that for any value of
magnetic field strength in the relativistic plasma, the upstream proper
speed of the plasma in the shock frame has to be at least greater
than 10 to attain the universal injection index. We require the
universal injection index to be obtained by all sources in order to
explain the observed similar injection index in two episodes of
DDRGs; otherwise, there has to be an adjustment of the parameter
values to produce similar observed injection indices in the two episodes
for our sample of DDRGs. For a jet head radius of 0.5 kpc, the spine
of the jet has to have $\Gamma_{\rm j}>50$ to produce the upstream jet
speed of $\Gamma_{\rm j,hs}>10$ for our study of the realistic case of
a hypothetical DDRG (see Fig.~\ref{fig_betah.vs.np_plot_3}).

The jet power of FR\,II sources spans 3 to 4 orders of magnitude
($\sim10^{43}-10^{47}$ erg s$^{-1}$), while it appears that $\Gamma_{\rm j}$ 
may vary only within a small range, as Mullin \& Hardcastle (2009) 
find no evidence for a correlation between bulk Lorentz factor and 
jet power. This means that higher jet power in
stronger sources cannot be produced solely by an increase in the jet
Lorentz factor. Instead, higher powers must be a result of higher
energy density in the jets, which implies higher energy densities in
the post-JTS jet matter. Since observations imply that hotspots are
close to equipartition (e.g. Hardcastle et al. 2004) this also implies
higher magnetic field strengths in the hotspots, with consequently
stronger synchrotron losses in the regions of particle acceleration.
This gives a way in which the jet power may affect the injection
index. Hardcastle \etal\ (2004) and Brunetti \etal\ (2003) argue,
based on X-ray and optical observations, that both the high-energy
cutoff and the break energy of the electron energy spectrum in the
hotspots are affected by the hotspot magnetic field strength. If the
hotspot break frequency is low enough that adiabatic expansion can
shift it down below the lowest frequency we observe (usually $\sim20$
MHz), then we expect to see an increased injection index, though we
would predict a spectral flattening at still lower frequencies. For a
break frequency around the optical/IR, we require one-dimensional
adiabatic expansion factors $\ga 30$ to bring the hotspot break
frequency down to hundreds of MHz, which is by no means implausible.

\begin{figure}
\hbox{
 \psfig{file=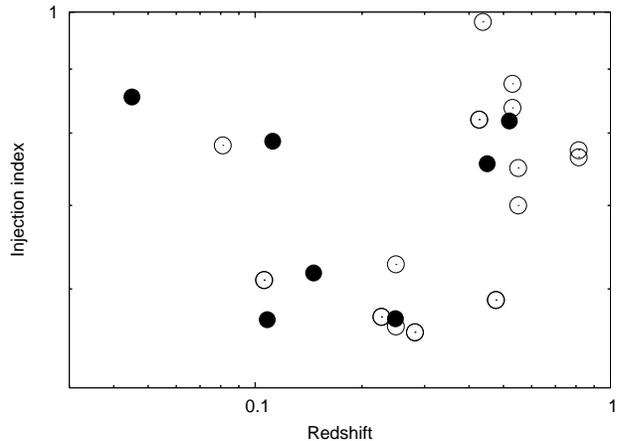,height=6.0cm,angle=-90}
}
\caption[]{$\alpha_{\rm inj}$ vs. $z$ plot for the same sample of sources as in 
           Fig.~\ref{fig_alpha-alpha_corrln}. Symbols have the same meaning as 
           in Fig.~\ref{fig_alpha-alpha_corrln}. No correlation is evident.}
\label{fig_alpha-z_no.corrln}
\end{figure}

The existence of correlations between the monochromatic radio
luminosity $L_{\nu}$, the source redshift $z$, and the low-frequency
spectral index $\alpha$ for radio galaxies have been known for some
time (Laing \& Peacock 1980). Since a higher $Q_{\rm j}$ produces higher
radio power in a source, our discussion above suggests that the
$\alpha-L_{\nu}$ correlation is the primary one, and is driven by our
proposed $\alpha_{\rm inj}-Q_{\rm j}$ correlation. In a flux-limited
sample $L_{\nu}$ and $z$ are correlated, so the $\alpha-z$ correlation
automatically arises. The fact that $\alpha-L_{\nu}$, or in a new
guise $\alpha_{\rm inj}-Q_{\rm j}$ is the primary correlation is
further supported by the absence of any correlation between
$\alpha_{\rm inj}$ and $z$ for the sample that we use in this paper
(see Fig.~\ref{fig_alpha-z_no.corrln}).

  We emphasise that the preceding discussion is based on the
  assumption that particle acceleration takes place predominantly at
  the JTS. As noted in Section 1, some particle acceleration certainly
  does occur in the pc- and kpc-scale jets, and while similar
  arguments to those given above may apply to the acceleration regions
  in the jets, we do not know enough about the acceleration mechanism
  for a detailed analysis. If the jets' initially ultrarelativistic
  speeds persist to large distances, as we suggest above, then it is
  hard to see how particle acceleration in the jet can have a
  significant effect on the post-JTS spectrum; moreover, for an
  ultrarelativistic shock, the state of the downstream plasma is
  essentially independent of that of the upstream plasma so long as
  the latter does not have a very high internal energy density\footnote{
  For the fast moving spine of the jet with moderate bulk Lorentz 
  factor of 4, we have shown in Appendix~\ref{sec_A4} that the internal 
  energy density is negligible compared to the bulk kinetic energy 
  density of the spine of the jet} (Kirk \etal\ 2000). While we 
  cannot rule out an effect of jet-related particle acceleration, 
  we can construct a self-consistent model in which it is unimportant.

\subsection{Testable predictions}
From our estimation, from equation~(\ref{eqn_tot.mass.accrn_2}) and
(\ref{eqn_Delta.a_2}), of total mass accumulation and change in spin
parameter $a$, for a central SMBH of with initial mass $\gapp10^8
M_{\odot}$, due to accretion can be significant if the SMBH accretes
efficiently over a time scales of a few $10^8$ yr or more. Though the
exact dependence is not known, the jet power is thought to depend on
$M_{\bullet}$ and $a$. Therefore, if the initial mass of a SMBH in any
radio galaxy is $\gapp10^8 M_{\odot}$ (all radio galaxies are supposed
to have $M_{\bullet} \gapp 10^8 M_{\odot}$: Chiaberge \& Marconi 2011)
and the duration of quiescent phase of jet activity is a few times $10^8$
yr or more, we might be able to find significantly different values of
injection spectral index in the two episodes. Since the fundamental
parameters that the jet power, $Q_{\rm j}$ is expected to depend on
are $M_{\bullet}$ and $a$, any change in these parameters by any means
should be in principle reflected in different values of the
injection index in two different episodes of a radio galaxy. In
misaligned DDRGs, the jet axes are changed significantly. Therefore, it
is quite natural to think that some large scale perturbation has
affected the central SMBH. Hence we are likely to see significantly
different values of injection index in two different episodes. As per
our expectation, we do see a significant difference in the values of
the injection index in two consecutive episodes of the misaligned
DDRG, 3C293. The error bars are smaller than the deviation from the
equality line. While this needs to be verified with a larger sample,
this is certainly consistent with our prediction. The error bars of
J0116-4722 (an aligned DDRG) are consistent with the injection indices
being similar in the two episodes.

From our detailed theoretical and observational studies we conjecture
that the DDRGs with $\Gamma_{\rm j} \lapp 10$ (if they exist) are
likely to have dissimilar injection index in two episodes, even though
$\Gamma_{\rm j}$ and $Q_{\rm j}$ are same in both episodes. The reason
is that $\Gamma_{\rm j, hs}$, and hence the Mach number of the JTS, will be
different in the two episodes, which leads to different injection indices.
Though Mullin \& Hardcastle (2009) suggest similar $\Gamma_{\rm j}$
for all FR\,II sources, they cannot completely rule out different
$\Gamma_{\rm j}$ for different sources. They look at the slowest outer
layer components of jet which may show up with the same/similar speed
even in the presence of variation in the Lorentz factors of jet
spines, hence a range of $\Gamma_{\rm j}$ values for the spines of the
jets might not be detected in their study. There could be very
low-power FR\,II sources which have low Lorentz factor ($<10$) spine
of their jets. If they are episodic then we predict that they should
show dissimilar injection indices in the two episodes.

  Finally, there is no real reason why DDRG with very different jet
  powers in the two episodes (due to a significant change in $\dot M$)
  should not exist in our model; they are just observationally
  difficult to detect since the inner and outer lobes would be
  expected to have very different luminosities/surface brightnesses.
  Sensitive low-frequency surveys with the GMRT and LOFAR might detect
  such objects in the future and our prediction is that large
  differences in the jet power in the two episodes will give rise to
  large differences in the measured injection indices.

\section{Summary and Conclusions}
\label{sec_conclusion}

We have discovered a new property of episodic (double-double) FRII
RGs, which allows us to draw many other inferences which may be 
summarized as follows. 
\begin{enumerate}
\item We have discovered that the values of injection index (the
  low-frequency power-law index of synchrotron emission) are
      similar in the two different episodes for most of the DDRGs 
      in our sample.
\item This observational result means that the injection index doesn't appear to bear any relation 
      to the medium into which the lobe is expanding. However, our
      sample (and previous work) show that it is strongly dependent on the jet power.
\item Point (ii) implies that the injection index over the 
      entire duration of the active phase of a radio galaxy 
      should be the same or similar provided the jet power remains constant, 
      even though the lobes are interacting with an ambient medium of
      strongly varying density.   
\item We have provided explicit, consistent relativistic formulae for
  the momentum balance at the jet head
  (equation~\ref{eqn_mom.balance}) and thus for the dynamics of the
  radio lobes in both episodes of the DDRG activity. We argue that
  observations favour $e^{-}e^{+}$ plasma rather
  than $e^{-}p^{+}$ plasma.
\item We argue that the jet power in the two
      episodes of a DDRG needs to be the same, or at least similar, with 
      the same or similar Lorentz factor and internal energy, in order
      to produce a similar injection index in both episodes
      without fine-tuning of other parameters governing the jet
      properties. An observational selection effect helps to enforce
      this condition in existing samples of DDRG.
\item Our models imply that the $\alpha-L_{\nu}$ correlation 
      of a flux limited sample of RGs is the primary one,
      and not the $\alpha-z$ correlation, and require that
      higher-power RGs do not have $\Gamma_{\rm j}$ but rather higher
      jet-frame energy density, consistent with observation. There may
      be a simple scaling relationship between the energy density in
      the jet and the accretion rate ${\dot M}$ for a radio-loud source..
\item We predict/conjecture that DDRGs with long quiescent phase, 
      misaligned doubles, low jet Lorentz factors ($<10$) or
      significant changes in jet power between the two episodes are more likely 
      to have dissimilar injection indices in two episodes of jet forming activity.
      These predictions can be tested observationally.

\end{enumerate}
\section*{Acknowledgments} 
This research has made use of the NASA/IPAC extragalactic database,
which is operated by the Jet Propulsion Laboratory, Caltech, under
contract with the National Aeronautics and Space Administration. We
thank the GNU/Linux group for their contribution. CK acknowledges Ronald
Taam for discussion on various theoretical issues. CK also thanks the
School of Physics, Astronomy and Mathematics at the University of
Hertfordshire, where a part of this research was done, for their
hospitality. CK acknowledges the grant (No. NSC99-2112-M-001-012-MY3)
from the National Science Council, Taiwan.

{}

\appendix
\section{Theoretical aspects of momentum balance}
\label{sec_A1}
\subsection{Momentum balance at the jet heads}
We consider a jet in the steady state (see Fig.~\ref{fig_jet.shocks}).
The momentum flow rate will remain conserved across the JTS as well as
the contact discontinuity. Since the jet fluid is dominated by
relativistic particles, it should be treated relativistically.
Hereafter, all the quantities we use are proper quantities unless
otherwise specified

We assume that the relativistic fluid description holds for the jet
matter. For the momentum balance of the relativistic flow of the
relativistic jet fluid at the hotspots, we start with the
energy-momentum tensor ($T^{ik}$) which has the form
\begin{equation}
T^{ik}=wu^iu^k-Pg^{ik},
\label{eqn_en.mom.tensor}
\end{equation}
(Landau \& Lifshitz 1966),
where $w$ is the enthalpy density; $u^i$ is the ith component of the 4-velocity of
the fluid, such that $u^0=\Gamma.1$ (temporal component) and $u^{\alpha}=\Gamma\beta^{\alpha}$ 
(spatial comonents, $\beta=\frac{v}{c}$ is not the proper velocity); 
and $g^{ik}$ is the metric tensor in flat space-time of the form
\begin{equation}
\begin{pmatrix}
1 & 0 & 0& 0 \\
0 &-1 & 0& 0 \\
0 & 0 &-1& 0 \\
0 & 0 & 0&-1 \\
\end{pmatrix}
\end{equation}
($i$, $k$, $\alpha$ and $0$ are not exponents, but indices to represents different components of the quantities).
\begin{equation}
w=e+P,
\end{equation}
where $e$ is the relativistic internal energy density, $P$ is the pressure of the jet fluid. 
In an explicit manner equation~(\ref{eqn_en.mom.tensor}) can be written as
\begin{equation}
\begin{pmatrix}
\Gamma^2e+\Gamma^2\beta^2P &\Gamma^2w\beta_x         & \Gamma^2w\beta_y          & \Gamma^2w\beta_z         \\
                           &                         &                           &                          \\
\Gamma^2w\beta_x           &\Gamma^2w\beta_x^2+P     & \Gamma^2w\beta_x\beta_y   & \Gamma^2w\beta_x\beta_z  \\
                           &                         &                           &                          \\
\Gamma^2w\beta_y           &\Gamma^2w\beta_y\beta_x  & \Gamma^2w\beta_y^2+P      & \Gamma^2w\beta_y\beta_z  \\
                           &                         &                           &                          \\
\Gamma^2w\beta_z           &\Gamma^2w\beta_z\beta_x  & \Gamma^2w\beta_z\beta_y   & \Gamma^2w\beta_z^2+P     \\
\end{pmatrix}.
\label{eqn_pmatrix}
\end{equation}
Since the radio galaxy jets are collimated jets, the fluid motion can very well be approximated as one dimensional
flow. Even though there must be velocity structure across the cross section of the jets, this is of no relevance 
to us, because the quantity of interest is the jet power ($Q_{\rm j}$) which is the time rate of flow of energy that passes through
a cross-sectional area held fixed at the host galaxy frame and normal
to the jet axis. Velocity structure, though it has been argued for in
some models of jet emission processes, has so far not been constrained
directly from observations. 

The spatial part of $T^{ij}$ will give us the momentum flow rate, which can be written as
\begin{equation} 
T^{\delta\mu} = \Gamma^2w\beta^{\delta}\beta^{\mu}+P\delta^{\delta\mu}
\label{eqn_en.mom.tensor_spatial}
\end{equation}
For one dimensional flow, say in the $x$ direction, we have to take
the projection 
of the tensor along the $x$ direction. The momentum flow rate ($\Pi$)
in the $x$ direction is given by
\begin{equation}
 \Pi(x)= <x\vert {\bf T} \vert x> = T^{xx},
\end{equation}
where $\vert x>$ is a unit bra-vector along the $x$ direction, and $<x \vert$ is its ket-vector. ${\bf T}$ is the 
energy-momentum  tensor.   
Since from equation~(\ref{eqn_pmatrix}),
\[ T^{xx} = \Gamma^2w(\beta)^2 +P \] (with $\beta_x$ replaced by $\beta$),
the momentum flow rate per unit area of jet cross section is as follows
\begin{equation}
 \Pi(x) = (\Gamma_{\rm j}^2w_{\rm j}\beta^2_{\rm j} +P_{\rm j})
\end{equation}
Following the treatment of Bicknell (1994), gravity, along with the
buoyancy force and any entrained momentum in the
jets can be neglected. We have assumed pressure-confined jets, and the jet pressure ($P_j$) and ambient medium pressure 
($P_a$) are negligible compared to jet momentum flux. So the net momentum flow rate per unit cross sectional area is
\begin{equation}
 \Pi_{\rm j}(x) = \Gamma_{\rm j}^2w_{\rm j}\beta^2_{\rm j}  
\end{equation}
This is the momentum flow rate in the host galaxy frame. 

In the hotspot frame (which is moving with a velocity of $\beta_{\rm
  hs}$ with respect to the host galaxy frame) the
above expression will have the same form but the quantities will be replaced by those in the hotspot frame
and can be written as $(\Gamma_{\rm j,hs}^2w_{\rm j}\beta^2_{\rm
  j,hs})A_{\rm j}$ (see Bicknell 1994 for a more 
regorous treatment). Thus, the momentum balance at the hotspot as seen in the hotspot frame can be written as
\begin{equation}
\Gamma_{\rm hs}^2w_{\rm a}\beta^2_{\rm hs}~A_{\rm h}  =  \Gamma_{\rm j,hs}^2w_{\rm j}\beta^2_{\rm j,hs}~A_{\rm j}
\label{eqn_mom.balance_1}
\end{equation}
We know that the the particles in the jets are accelerated in the
hotspots and thereby inflate lobes whose internal pressure is similar
to the external (or ambient medium) pressure (Croston et al. 2004;
Konar et al., 2009; Shelton, Kwak \& Henley, 2012). In relativistic
dynamics, the enthalpy contains a term accounting for the rest
mass energy. We can thus write $w$, for both the jet fluid and
ambient medium fluid, as
\begin{equation}
w= \rho^0 c^2 + \epsilon + P = \rho c^2 + P,
\label{eqn_enthalpy}
\end{equation}
where $\rho$ is the proper density of the magnetised thermal/nonthermal matter and can be expressed as
\begin{equation}
\rho = \rho^0 + \frac{\epsilon}{c^2}. 
\label{eqn_prop.density}
\end{equation}
Relating the quantities in the host galaxy frame and the hotspot frame
using velocity addition theory it can be shown that
\begin{equation}
\Gamma^2_{\rm j,hs}\beta^2_{\rm j,hs}=\Gamma^2_{\rm j}\Gamma^2_{\rm hs}(\beta_{\rm j}-\beta_{\rm hs})^2. 
\label{eqn_velo.addition}
\end{equation}
The jet power ($Q_{\rm j}$) as measured in the host galaxy frame can
be expressed as
\begin{equation}
Q_{\rm j}= \Gamma^2_{\rm j}w_{\rm j}(\beta_{\rm j}c)A_{\rm j}, 
\label{eqn_jetpow}
\end{equation}
where $Q_{\rm j}$ includes the rest mass energy of the jet matter.   

Replacing $\Gamma^2_{\rm j,hs}\beta^2_{\rm j,hs}$ from equation~(\ref{eqn_velo.addition}) into 
equation~(\ref{eqn_mom.balance_1}) and then replacing $\Gamma^2_{\rm j}w_{\rm j}A_{\rm j}$ from equation~(\ref{eqn_jetpow}),
the momentum balance equation becomes
\begin{equation}
\frac{Q_{\rm j}}{\beta_{\rm j}c} = w_{\rm a}A_{\rm h}(\frac{\beta_{\rm hs}}{\beta_{\rm j}-\beta_{\rm hs}})^2.
\end{equation}
This can be rearranged to give
\begin{equation}
\beta_{\rm hs} = \left(\frac{1}{1+\eta}\right)\beta_{\rm j},
\label{eqn_mom.balance_2}
\end{equation}
where 
\begin{equation}
\eta=\sqrt{\left(\frac{\beta_{\rm j}cA_{\rm h}w_{\rm a}}{Q_{\rm j}}\right)} 
\label{eqn_eta}
\end{equation}

\section{Algebraic expressions of some parameters in the expressions of fast and slow waves}
\label{sec_A2}

According to Chou \& Hao (2004), the fast and slow waves in a relativistic magnetoplasma, 
which exerts isotropic pressure, can be given by 
\begin{equation}
V_{\rm f,s}^2 = \frac{-b_{\rm is}\pm \sqrt{b_{\rm is}^2-4a_{\rm is}C_{\rm is}}}{2a_{\rm is}},
\end{equation}
where $a_{\rm is}$, $b_{\rm is}$ and $C_{\rm is}$ are given below.
\[a_{\rm is}= 1+\frac{1}{\rho^0c^2}\left(\frac{B^2}{4\pi}+ 2P^{\rm nt}+2\epsilon^{\rm nt}\right) + ~~~~~~~~~~~~~~~~~~~~~~~~~~~~~~~~~~~~\]
\begin{equation}
 ~~\frac{1}{(\rho^0c^2)^2}\left\{(\epsilon^{\rm nt}+P^{\rm nt})^2 + (\epsilon^{\rm nt}+P^{\rm nt})\frac{B^2}{4\pi}\right\},
\end{equation}
\[ b_{\rm is}= \frac{1}{\rho^0} (\gamma_{\rm ad}P^{\rm nt}+\frac{B^2}{4\pi}) + ~~~~~~~~~~~~~~~~~~~~~~~~~~~~~~~~~~~~~~~~~~~~~~~~~~~~~~~~~~ \]
\begin{equation}
       ~~~\frac{1}{{\rho^0}^2c^2} \left \{(\gamma_{\rm ad}P^{\rm nt}+\frac{B^2}{4\pi})(\epsilon^{\rm nt}+P^{\rm nt}) 
             +  (\gamma_{\rm ad}P^{\rm nt})(\frac{B^2}{4\pi})\cos^2\theta \right\}  
\end{equation}
and
\begin{equation}
C_{\rm is} = \frac{\gamma_{\rm ad}P^{\rm nt}}{{\rho^0}^2}\frac{B^2}{4\pi}\cos^2\theta. ~~~~~~~~~~~~~~~~~~~~~~~~~~~~~~~~~~~~~~~~~~~~~~~~~~~~~~~~~~~~~~~~~~~~~~~~~
\end{equation}
In the limit $\rho^0c^2 \rightarrow 0$, $a_{\rm is}$ and $b_{\rm is}$ blow up.
Since we have to deal with $\rho^0c^2 \rightarrow 0$ limit, we better work with $b^{'}$ and $C^{'}$,
where
\[b^{'}= \frac{b_{\rm is}}{a_{\rm is}}=~~~~~~~~~~~~~~~~~~~~~~~~~~~~~~~~~~~~~~~~~~~~~~~~~~~~~~~~~~~~~~~~~~~~~~~~~~ \]
{\scriptsize
\begin{equation}
\frac{(\gamma_{\rm ad}P^{\rm nt}+\frac{B^2}{4\pi})\rho^0c^2 + (\gamma_{\rm ad}P^{\rm nt}+\frac{B^2}{4\pi})(\epsilon^{\rm nt}+P^{\rm nt}) 
                               + (\gamma_{\rm ad}P^{\rm nt})(\frac{B^2}{4\pi})\cos^2\theta}
{(\rho^0c^2)^2 + (\frac{B^2}{4\pi}+ 2P^{\rm nt}+2\epsilon^{\rm nt})\rho^0c^2 + (\epsilon^{\rm nt}+P^{\rm nt})(\epsilon^{\rm nt}+P^{\rm nt}+\frac{B^2}{4\pi})}c^2
\end{equation}
}
and
\[C^{'}= \frac{C_{\rm is}}{a_{\rm is}}=~~~~~~~~~~~~~~~~~~~~~~~~~~~~~~~~~~~~~~~~~~~~~~~~~~~~~~~~~~~~~~~~~~~~~~~~ \]
{\scriptsize
\begin{equation}
\frac{(\gamma_{\rm ad}P^{\rm nt})(\frac{B^2}{4\pi})\cos^2\theta}{(\rho^0c^2)^2 + (\frac{B^2}{4\pi}+ 2P^{\rm nt}+2\epsilon^{\rm nt})\rho^0c^2 + (\epsilon^{\rm nt}+P^{\rm nt})(\epsilon^{\rm nt}+P^{\rm nt}+\frac{B^2}{4\pi})}c^4.
\end{equation}
}
In the limit $\rho^0c^2 \rightarrow 0$, even if $a_{\rm is}$ and $b_{\rm is}$ blows up, $b^{'}$ and $C^{'}$
does not. Therefore, just replacing $\rho^0c^2$ by zero, we will get the expression for $V_{\rm I}$, $V_{\rm f}$ and $V_{\rm s}$
in the ultrarelativistic limit.

\section{Estimation of mass and spin accumulation of supermassive black hole during quiscent phase}
\label{sec_A3}
\subsection{Mass accumulation}
If we assume that the SMBH accretes and radiates at a constant fraction of 
the Eddington rate in the quiescent phase, we can write
\begin{equation}
L_{\rm bol}=\lambda L_{\rm Edd},
\label{eqn_accrn.lum}
\end{equation}
where $L_{\rm bol}$ is bolometric luminosity and $L_{\rm Edd}$ is the Eddington 
luminosity. If the radiative efficiency factor is $\xi$, then the mass accretion 
rate is
\begin{equation}
\dot{M} = \frac{L_{\rm bol}}{\xi c^2}.
\label{eqn_accrn.rate}
\end{equation}
Application of equations~(\ref{eqn_accrn.lum}) and (\ref{eqn_accrn.rate}) yields
\begin{equation}
\Delta M = \int_0^{\Delta t} \dot{M} dt = \frac{\lambda}{\xi} \frac{L_{\rm Edd}}{c^2} \Delta t
\label{eqn_tot.mass.accrn_1}
\end{equation}
Expressing in astronomical units, equation~(\ref{eqn_tot.mass.accrn_1}) becomes
\begin{equation}
\left(\frac{\Delta M}{M_{\odot}}\right)  = 2.27\times 10^{-9} \left[ \frac{\lambda}{\xi} \left(\frac{M_{\bullet}}{M_{\odot}}\right) \left (\frac{\Delta t}{\rm yr} \right) \right],
\label{eqn_tot.mass.accrn_2}
\end{equation}
where $M_{\odot}$ is solar mass.

\subsection{Spin accumulation}
We use the Newtonian physics for our crude estimate just to have an
idea about the situation. The specific orbital angular momentum of the
disk matter at ISCO is given by
\begin{equation}
J=V_k3R_s = \sqrt{GM_{\bullet}3R_s} = \sqrt{6}\frac{GM_{\bullet}}{c},      
\end{equation}
where $V_k$ is the Keplerian speed (substituting in for $R_s$).
When the matter is accreted onto the black hole, the orbital angular momentum of
the matter adds to the spin angular momentum, $S_{\bullet}$ of the black hole.
The total change in angular momentum, $\Lambda$, over the time $\Delta t$ can
be expressed as
\[\Delta \Lambda = \pm J\times \Delta M,  \]
where $+$ and $-$ sign are for a co-rotating and counter-rotating disk
with respect to the black hole spin.
Substituting the expression of $\Delta M$ from equation~(\ref{eqn_tot.mass.accrn_1}),
we obtain
\begin{equation}
\Delta \Lambda = \pm \frac{\sqrt{6}GM_{\bullet}}{c}\times \frac{\lambda}{\xi} \frac{L_{\rm Edd}}{c^2} \Delta t 
\label{eqn_Delta.Lambda}
\end{equation}
We can calculate the increase in the dimensionless angular momentum parameter, $a$ which is
given by
\begin{equation}
a=\frac{S_{\bullet}c}{GM^2_{\bullet}}.
\end{equation}
The change in $a$ is the total orbital angular momentum (in dimensionless units) of the
accreted matter, and is given by
\[ \Delta a = \pm \frac{(\Delta \Lambda) c}{GM^2_{\bullet}}. \]
Substituting $\Delta \Lambda$ from equation~(\ref{eqn_Delta.Lambda}), we obtain
\begin{equation}
\Delta a = \pm \sqrt{6}\frac{\lambda}{\xi}\frac{L_{\rm Edd}}{M_{\bullet}c^2} (\Delta t)
\end{equation}
Replacing $L_{\rm Edd} = 1.3\times 10^{38}(\frac{M_{\bullet}}{M_{\odot}})$ erg s$^{-1}$
and expressing $\Delta t$ in yr,  the above expression can be written as
\begin{equation}
\Delta a = \pm \sqrt{6}\frac{\lambda}{\xi}\frac{4.1\times 10^{45}}{M_{\odot}c^2} (\frac{\Delta t}{\rm yr})
\label{eqn_Delta.a_1}
\end{equation}
This expression is independent of $M_{\bullet}$. With the same canonical values of
$\lambda$ and $\xi$, for the duration of quiescent phase, $\Delta t = 10^7$ yr,
\begin{equation}
\Delta a = \pm 0.056  
\label{ean_Delta.a}
\end{equation}
Since, we have assumed that the ISCO is at $3R_s$, we can use pseudo General Relativistic (GR)
potential (Paczy{\'n}sky \& Wiita 1980) to estimate values of $V_k$ to get the better estimation
of $\Delta a$. Equating gravitational force with the centrifugal force we get the following
expression of the specific angular momentum:
\begin{equation} 
J =V_k r = \sqrt{GM_{\bullet}r}\left(\frac{r}{r-R_s}\right).
\end{equation}
Replacing $r=3R_s$, we get
\begin{equation} 
J = \sqrt{GM_{\bullet}3R_s}\frac{3}{2}  
\end{equation}
In the validity regime of the pseudo-GR potential, the Newtonian expression of $\Delta a$
(equation~\ref{eqn_Delta.a_1}) should be multiplied by a factor of $\frac{3}{2}$.
The GR-corrected expression should therefore be
\begin{equation}
\Delta a = \pm \sqrt{6}\frac{\lambda}{\xi}\frac{6.15\times 10^{45}}{M_{\odot}c^2} (\frac{\Delta t}{\rm yr}).
\label{eqn_Delta.a_2}
\end{equation}
This gives us
\begin{equation}
\Delta a = 0.056\times \frac{3}{2} = 0.084. 
\end{equation}


\section{Ratio of internal energy density to kinetic energy density of the jet fluid in the spine of the jet}
\label{sec_A4}
The kinetic energy ($e_{\rm ke,j}$) content per unit proper volume of the jet is given by
\[ e_{\rm ke,j} =\Gamma_{\rm j}(\Gamma_{\rm j}-1)\rho^0_{\rm j}c^2 + \Gamma_{\rm j}(\Gamma_{\rm j}-1)(\epsilon_{\rm j} + P_{\rm j}) \]
\[~~~~= \Gamma_{\rm j}(\Gamma_{\rm j}-1) [ \rho^0_{\rm j}c^2 + \gamma \epsilon_{\rm j}], \] 
we have used the relation $P_{\rm j} = (\gamma -1)\epsilon_{\rm j}$, where $\gamma$ is the 
adiabatic index and $\epsilon_{\rm j}$ is the internal energy (without the rest mass energy) density 
of the jet fluid.  With a little algebra, we obtain
\begin{equation}
\epsilon_{\rm j} = \frac{1}{\Gamma_{\rm j}(\Gamma_{\rm j} -1) \gamma \left[1+\frac{1}{\gamma}\chi_{\rm j}\right] }e_{\rm ke,j}, 
\label{eqn_ej.by.ek}
\end{equation} 
where $\chi_{\rm j}=\frac{\rho^0_{\rm j}c^2}{\epsilon_{\rm j}}$.  

 Whatever the EoS of the jet fluid may be, the adiabatic index
  $\gamma$ must obey the inequality $\frac{4}{3} \le \gamma \le
  \frac{5}{3}$. So $\gamma$ is always $> 1$. Therefore, whatever the value of
  $\chi_{\rm j}$ ($\chi_{\rm j}>0$) for FR\,II jets, the quantity
  within the bracket in the denominator of equation~(\ref{eqn_ej.by.ek})
  is greater than 1. So, for any value of $\Gamma_{\rm j}>1$, $ \epsilon_{\rm
    j} < e_{\rm ke,j}$. Now if we are to believe X-ray inverse-Compton
  studies (e.g. Hardcastle 2006) that
  argue that $\Gamma_{\rm j} > 15$ for the spine of the jet, then we
  can safely conclude that $\epsilon_{\rm j} \ll e_{\rm ke,j}$, and hence
  the internal energy content of the pre JTS jet fluid is negligible
  compared to the bulk kinetic energy of the jet fluid. Even for
  a more modest value of $\Gamma_{\rm j}=4$, which is not an absurd value for the
  jet spine, $ \epsilon_{\rm j}$ is lower than $e_{\rm ke,j}$ by a
  factor that is greater than $\sim 12$. 
\label{lastpage}
\end{document}